\documentclass{aa}

\usepackage{graphicx}	
\usepackage{amsmath, physics}	
\usepackage{booktabs, float, makecell, array} 
\usepackage{siunitx}   
\usepackage{textcomp}  
\usepackage{gensymb}   
\usepackage{placeins}
\usepackage{caption}      
\usepackage{subcaption}   
\usepackage{multirow}     

\usepackage{txfonts}
\usepackage{hyperref}
\hypersetup{
    colorlinks=true,
    citecolor=blue,
    linkcolor=red,
    filecolor=magenta,
    urlcolor=blue,
}

\newcommand{\dash}{\text{--}}
\newcommand{\LT}{L_\mathrm{X}\dash T}
\newcommand{\YT}{Y_\mathrm{SZ}\dash T}
\newcommand{\MT}{M_\mathrm{gas}\dash T}

\bibpunct{(}{)}{;}{a}{}{,} 

\begin{document}

\title{Characterising galaxy cluster scaling relations as cosmic isotropy tracers using FLAMINGO simulations}

\author{Yujie He\thanks{\email{yujiehe@strw.leidenuniv.nl}}\inst{1}
          \and
        Konstantinos Migkas\inst{1,2}
          \and
        Joop Schaye\inst{1}
          \and
        Joey Braspenning\inst{3,1}
          \and
        Matthieu Schaller\inst{4,1}
}

\institute{$^1$ Leiden Observatory, Leiden University, PO Box 9513, 2300 RA Leiden, the Netherlands  \\ 
$^2$SRON Netherlands Institute for Space Research, Niels Bohrweg 4, NL-2333 CA Leiden, the Netherlands\\
$^3$Max Planck Institute for Astronomy, Königstuhl 17, D-69117 Heidelberg, Germany\\
$^4$Lorentz Institute for Theoretical Physics, Leiden University, PO box 9506, 2300 RA Leiden, the Netherlands\\
}

   \date{}


    \abstract{
    The standard cosmological model, $\Lambda$CDM, assumes isotropy on large cosmic scales. However, recent studies using galaxy cluster scaling relations have reported an apparent $H_0$ anisotropy at $5.4\sigma$ that could be attributed to large bulk flows extending beyond $\SI{500}{Mpc}$, which is in disagreement with $\Lambda$CDM. To quantify the statistical tension of the observational galaxy cluster data used in past studies with $\Lambda$CDM, we utilised the isotropic ($\SI{2.8}{Gpc})^3$ run of the FLAMINGO ($\Lambda$CDM) simulations, the largest hydrodynamical cosmological simulation available to date. We created 1728 simulated lightcones and studied the apparent level of anisotropy traced by X-ray and thermal Sunyaev-Zeldovich scaling relations in the same cluster sample selection and methodology as in the past study. We find the probability of such apparent anisotropies randomly emerging in cluster scaling relations within a $\Lambda$CDM universe to be $0.12\%\, (3.2\sigma)$. The discrepancy goes up to $\sim 3.6\sigma$ when modelled as a bulk flow at $z < 0.1$. We also find that statistical noise accounts for over $80\%$ of the anisotropy amplitude in each lightcone, with large peculiar velocities contributing less than $20\%$. We also show that anisotropy amplitudes are highly sensitive to the intrinsic scatter in the scaling relations, with tighter relations providing stronger constraints. Nevertheless, the tension between the past results and $\Lambda$CDM persists, albeit at a lower significance than previously reported.}


   \maketitle
%

\section{Introduction}

One of the most important underlying assumptions of our standard model of cosmology, $\Lambda$CDM, is the statistical homogeneity and isotropy of space and matter. This is known as the cosmological principle (CP). This model has been remarkably predictive in explaining various cosmic phenomena, including the power spectrum of the cosmic microwave background (CMB), the abundance of light elements, and the distribution of large-scale structures. However, recent observations have revealed several notable tensions with $\Lambda$CDM, including the long-standing $H_0$ tension \citep[e.g.][]{riess_large_2019, riess_comprehensive_2022, planck_collaboration_planck_2020, valentino_realm_2021, shah_buyers_2021} and the $S_8$ tension \citep[e.g.][]{joudaki_cfhtlens_2017, des_collaboration_dark_2022, philcox_boss_2022, mccarthy_flamingo_2023, ghirardini_srgerosita_2024}. In addition, many studies have examined the validity of CP, and a number of anomalies have been reported, including the cosmic dipole \citep{rubart_cosmic_2013, tiwari_dipole_2015, secrest_test_2021, secrest_challenge_2022, dam_testing_2023, oayda_testing_2023, mittal_cosmic_2024,da_silveira_ferreira_tomographic_2024,tiwari_independent_2024}, SNIa anisotropy \citep{chang_comparison_2015, mohayaee_supernovae_2021, cowell_potential_2023, hu_testing_2024, perivolaropoulos_isotropy_2023,mc_conville_anisotropic_2023}, excessive bulk flows on large scales\footnote{{Some studies do not claim a strong tension with $\Lambda$CDM but find `atypical' bulk flows \citep[e.g.][$3.4\sigma$ at $250h^{-1}\SI{}{Mpc}$]{hoffman_large-scale_2024} or external bulk flows that do not converge on larger scales \citep[][Fig.~9]{carrick_cosmological_2015}. Bulk flow studies tracing smaller volumes are more often consistent with $\Lambda$CDM \citep[e.g.][$<100h^{-1}\SI{}{Mpc}$]{nusser_cosmological_2011}.}} \citep{kashlinsky_new_2010,  salehi_cosmological_2021, whitford_evaluating_2023, watkins_analysing_2023}, and last but not least, anisotropic Hubble expansion \citep{migkas_cosmological_2021, luongo_larger_2022, pandya_examining_2024, haridasu_radial_2024,boubel_testing_2025}. We refer to some reviews for more on cosmic tensions and anomalies \citep{bull_beyond_2016, abdalla_cosmology_2022, perivolaropoulos_challenges_2022, aluri_is_2023}. {While there are indications of discrepancies, no definitive answer or consensus has been reached on whether the anomalies are real tensions or statistical artefacts.}

The CP provides the foundation of our cosmology theory and a cosmic reference frame for observations. The importance of robust tests of the CP cannot be overstated. As we enter an era of high-precision multi-messenger cosmology, with a wide range of experiments to probe local and high-redshift Universe, it is crucial to fully understand the possible systematic biases that creep into the past results and to quantify their true tension with $\Lambda$CDM.

The $\Lambda$CDM adoption of the CP posits that the matter rest frame should converge to the CMB rest frame {\citep[e.g.][]{ferreira_first_2021,wagenveld_meerkat_2024}} at scales $\gtrsim \SI{150}{Mpc}$\footnote{This scale is decided by the cosmological model and not the CP itself.} \citep{fixsen_cosmic_1996, planck_collaboration_planck_2014}. On such scales, the Hubble flow is expected to be statistically uniform across the sky, with negligible bulk flow motions {\citep[$\lesssim \SI{140}{km.s^{-1}}$, e.g.][]{watkins_analysing_2023}}, as dictated by the statistical homogeneity of the cosmic matter distribution. However, if such bulk flows are ignored, they can result in an observed $\lesssim 3\%$ anisotropy in the redshift-distance relation and the Hubble constant ($H_0$) within the CMB rest frame \citep{migkas_galaxy_2025}. However, recent studies suggest that the galaxy cluster rest frame does not converge to the CMB rest frame out to $z\sim 0.1\dash 0.2$ \citep{migkas_anisotropy_2018, migkas_probing_2020, migkas_cosmological_2021,pandya_examining_2024}. (For a detailed review, see \citet{migkas_galaxy_2025}).

Galaxy clusters are the largest virialised structures in the Universe. These massive objects are typically observed at redshift $z\sim0.05\dash 0.5$ with a uniform distribution across the sky, rendering them a powerful cosmology probe into the late Universe. \citet[][hereafter M21]{migkas_cosmological_2021} studied the cosmic (an)isotropy within $\lesssim 1$ Gpc using X-ray, microwave, and infrared data for an X-ray flux-limited cluster sample with 313 galaxy clusters introduced in \citet[][hereafter M20]{migkas_probing_2020}. They also utilised the ASCA cluster sample \citep{horner_x-ray_2001} containing another $\sim 160$ independent objects. They constructed the multi-wavelength scaling relations $\LT$, $\YT$, and $L_\mathrm{BCG}\dash T$, where $L_\mathrm{X}$ is the X-ray luminosity, $Y_\mathrm{SZ}$ is the total integrated Compton parameter, $L_\mathrm{BCG}$ is the near-infrared cluster central galaxy luminosity, and $T$ is the cluster X-ray temperature. While the determination of $L_\mathrm{X}$, $Y_\mathrm{SZ}$, and $L_\mathrm{BCG}$ depend strongly on the underlying cosmology, $T$ is measured independently of any cosmological assumptions. Consequently, angular variations of cosmological parameters and cluster bulk flow motions are imprinted in the directional behaviour of these scaling relations. Exploiting this feature, M21 detected an apparent 9\% dipole variation of $H_0$ pointing towards $(l,b)= (280\degree\pm{35\degree}, -15\degree\pm{20\degree})$ at $5.4\sigma$. Alternatively, they showed that their findings could be attributed to a bulk flow of $\sim \SI{900}{km.s^{-1}}$ towards the opposite direction that extends to $\gtrsim \SI{500}{Mpc}$. Despite very different potential systematics relevant to each relation and sample, the apparent anisotropy was consistently detected in all three nearly independent scaling relations and the two cluster samples. An extensive set of tests was performed to ensure that the result did not originate from any known systematic biases. More recently, \citet{pandya_examining_2024} used cluster velocity dispersion ($\sigma_v$) data and the $L_\mathrm{X}\dash \sigma_v$ and $Y_\mathrm{SZ}\dash\sigma_v$ scaling relations to probe cosmic isotropy. They detected a $(27.6\pm4.4\%)$ $H_0$ dipole\footnote{\citet{pandya_examining_2024} have also reported that the level of their observed $H_0$ variation is overestimated due to the large scatter of the used scaling relations. They estimate that the `true' $H_0$ dipole they detect on top of statistical noise is $\sim 8\%$.} pointing towards $(295\degree\pm 71\degree , -30\degree \pm 71 \degree)$ at a $3.6\sigma$ level, which is consistent with M21. The existence of such a dipole is recently confirmed independently by \citet{boubel_testing_2025} using the galaxy Tully-Fisher relation, which finds a $3\pm 0.7\%$ $H_0$ variation in $(l,b)=(322\pm30\degree, -52\pm 10\degree)$, close to M21 and \citet{pandya_examining_2024}. We note that we report the lowest $H_0$ directions for these works.

Despite the assumed isotropy on large scales in $\Lambda$CDM, one may observe that anisotropies emerge simply due to cosmic and sample variance as well as from residual bulk flow motions. The amplitude and significance of such observed anisotropies could be further upscattered due to statistical noise and systematic biases in the data selection and applied statistics. Furthermore, one needs to accurately know the expected observed $H_0$ variation within a $\Lambda$CDM universe to properly assess the tension between the {observational} findings {in M21} and the $\Lambda$CDM model. Arguably the most optimal way to achieve all of the above is to apply the same sample selection and statistical techniques {of M21} to isotropic cosmological hydrodynamical simulations.

The Virgo Consortium's FLAMINGO (Full-hydro Large-scale structure simulations with All-sky Mapping for the Interpretation of Next Generation Observations) simulations include the largest state-of-the-art hydrodynamical cosmological simulation to date. FLAMINGO is described in detail in \cite{schaye_flamingo_2023}, hereafter S23. These simulations are designed for studies of galaxy clusters and other large-scale structures. We used its flagship run, the $(\SI{2.8}{Gpc})^3$ simulation box using a baryon particle mass of $1.1\times 10^9 \,\mathrm{M}_\odot$ mass resolution (named L2p8\_m9), reaching $z = 0$ with $2.8\times 10^{11}$ particles. The $(\SI{1.0}{Gpc})^3$ fiducial run (named L1\_m9), which uses the same model and resolution but only a smaller box and lower particle count of $1.3\times 10^{10}$, is sometimes used for comparison purposes. The FLAMINGO project incorporates rich baryonic physics and covers an extensive cosmic volume, rendering it an excellent laboratory for our purpose. Additionally, the cluster thermodynamic profiles \citep{braspenning_flamingo_2024} and scaling relations (S23) have been shown to adequately match observations.

In this work, we aim to quantify the probability that the anisotropies observed in M21 arise in a $\Lambda$CDM universe. To this end, we placed 1728 observers within the FLAMINGO simulation volume of $(\SI{2.8}{Gpc})^3$ and constructed 1728 isotropic lightcone samples. We applied the same cluster selection criteria and statistical procedures as in M21, and we injected realistic scatter and measurement uncertainties into the simulated cluster catalogues. This enabled us to assess the true level of tension between the multi-wavelength galaxy cluster observations and the $\Lambda$CDM model, while accounting for potential systematics in data selection, anisotropy modelling, statistical methodology, and random noise. In addition to the M21 framework, we employed an Markov chain Monte Carlo (MCMC)  based statistical approach for cross-validation. We further investigated systematics that may artificially inflate the amplitude or significance of the observed anisotropies {in M21}, such as peculiar velocities and scatter in scaling relations. This study is an essential step towards establishing cluster scaling relations as a robust probe of $H_0$ variation and bulk flows.

This paper is organised as follows: In Sect.~\ref{sec:sample}, we describe the procedures used to obtain the lightcones and the cluster catalogues. In Sect.~\ref{sec:method}, we introduce the statistical methods employed to constrain anisotropy and bulk flows. In Sect.~\ref{sec:results}, we present the main results and assess the probability at which the M21 result reconciles with $\Lambda$CDM. In Sect.~\ref{sec:discussion}, we discuss the implications of our findings and the systematics that could affect our analysis. Finally, we conclude in Sect.~\ref{sec:conclusion}.

We note that in calculating distances and the evolution factor, $E(z)$ (Sect.~\ref{sec:method}), we followed the Dark Energy Survey year three cosmology \citep[$H_0=\SI{68.1}{km.s^{-1}.Mpc^{-1}}$, $\Omega_\mathrm{m}=0.306$, $\Omega_\Lambda=0.694$,][]{des_collaboration_dark_2022}, which the FLAMINGO fiducial runs (L1\_m9, L2p8\_m9) we used are based on. Under all circumstances, we modelled only the relative difference of $H_0$, and the fiducial choice of $H_0$ therefore has no bearing on the result.

\section{FLAMINGO lightcone generation and cluster sample construction}\label{sec:sample}

Typical numerical simulations, such as FLAMINGO, output snapshots of the simulation box at fixed cosmic times. However, observations are limited to an observer's past lightcone, where objects are seen as they were when the light they emitted first reached the observer. To replicate this in simulations, mock lightcones are constructed by stitching together multiple snapshots and assigning each object a redshift and sky position based on the time it crossed the observer's past lightcone. This process assumes a cosmological model and produces a lightcone catalogue that accounts for the evolution of structures over cosmic time, enabling direct comparison with observations.

The identification of galaxy clusters in FLAMINGO was performed using the VELOCIraptor subhalo finder \citep{elahi_hunting_2019}, which implemented a Friends-of-Friends (FOF) algorithm. After this, the halo properties were computed using the Spherical Overdensity and Aperture Processor (SOAP)\footnote{ \url{https://github.com/SWIFTSIM/SOAP}} {\citep{mcgibbon_soap_2025}}, a tool developed by the FLAMINGO team alongside the simulations to compute various (sub)halo properties (McGibbon, in prep.). We used the SOAP catalogue to obtain $L_\mathrm{X}$, $Y_\mathrm{SZ}$, $T$, the cluster gas mass $M_{\text{gas}}$, and the centre-of-mass velocity $\vb{v}$.

FLAMINGO produced eight lightcones at runtime for the $(\SI{2.8}{Gpc})^3$ run. In runtime lightcones, the particle position between snapshots is available, allowing for a more accurate estimation of the particle location $\vb{x}$ and redshift $z$ at the moment of crossing the lightcone (S23). Despite the high quality of these data, eight lightcones are not sufficient to robustly assess the rarity of the M21 results. Consequently, through a post-processing procedure, we created 1728 evenly spaced lightcones, 12 on each side, in the $(\SI{2.8}{Gpc})^3$ simulation box.

\subsection{FLAMINGO lightcones construction}\label{sec:lightcone-construction}

The observers were placed at fixed comoving coordinates $(iL/12,\,jL/12,\,kL/12)$, where $i,j,k$ are integers from $-6$ to $+5$ and $L=\SI{2.8}{Gpc}$ is the box size. For consistency with M21, we considered only clusters at $z\leq 0.30$, corresponding to a comoving distance of $\sim\SI{1.2}{Gpc}$. Given SOAP's cadence $\Delta z=0.10$, four snapshots are required to construct the lightcones to $z\leq 0.30$. This redshift range is well below the half-box size $\SI{1.4}{Gpc}$, avoiding edge effects and box repetition. For our study, box repetition could introduce false statistics by duplicating the same scaling behaviour, biasing the scaling relation fit. Additionally, $z<0.30$ is sufficient to include most of the flux-selected clusters, with a higher-redshift tail containing less than 1\% of the sample, as seen in M20 and M21.

Lightcones were created as follows: for each halo in snapshot $n$, at cosmic redshift $z_n$, we calculated the comoving distance $r$ from the cluster to the observer and compare it with $r(z_n) = \int_0^{z_n}\frac{c\dd{z}'}{H(z')}$. The cluster is included in the lightcone if its comoving distance falls within the shell $r(z_n - \Delta z/2) < r < r(z_n + \Delta z/2)$. We then solved for the precise cosmic redshift, $z_\mathrm{cos}$, that satisfies the relation between comoving distance and light travel distance along radial null geodesics:
\begin{align}\label{eq:lightcone}
    - \int_0^{z_\mathrm{cos}} \frac{c\dd{z'}}{H(z')} + r(a) = 0.
\end{align}
After $z_\mathrm{cos}$ was found, all properties of the halo, including coordinates and velocities, were taken from snapshot $n$, for which $z_n$ is closest to $z_\mathrm{cos}$. To remove the galaxy groups and reduce the computational overhead, we applied a mass selection of $M_\mathrm{500c} > 10^{13} \,\mathrm{M}_\odot$,\footnote{Clusters or galaxy groups with $M_\mathrm{500c} < 10^{13}\,\mathrm{M}_\odot$ rarely pass our flux selection (see Appendix~\ref{app:mass-cut}).} where $M_\mathrm{500c}$ is the mass enclosed within a region of density 500 times the critical density of the Universe at the cluster redshift.

Once $z_\mathrm{cos}$ was determined, it was combined with the Doppler redshift (due to the peculiar velocity), $z_\mathrm{pec}$, to compute the observed redshift, $z_\mathrm{obs}$:
\begin{align}
1 + z_\mathrm{obs} = (1 + z_\mathrm{cos})(1 + z_\mathrm{pec}). \label{eq:zobs}
\end{align}
In our case, $z_\mathrm{obs}$ is equivalent to the CMB rest-frame redshift in observations, affected only by cosmic expansion and the peculiar velocity of the source, and not by any peculiar motion of the observer within the cosmic rest frame. The relativistic Doppler effect was computed as
\begin{align}
1 + z_\mathrm{pec} = \gamma[1 + v_\mathrm{pec} \cos\theta / c],
\end{align}
where $\theta$ is the angle between the cluster's peculiar motion and its line-of-sight to the observer, $v_\mathrm{pec}$ is the total peculiar velocity, $\gamma = [1 - (v_\mathrm{pec}/c)^2]^{-1/2}$, and $c$ is the speed of light. The observed redshift is ultimately used to model bulk flow and $H_0$ variation.

We used the SOAP properties computed at time steps $\Delta z=0.10$ and did not perform any interpolation between snapshots. The general accuracy of phase-space interpolation of cluster velocities and positions in $N$-body simulations remains to be determined. The decision not to interpolate was previously adopted for galaxy lightcones \citep[e.g.][]{kitzbichler_high-redshift_2007}, where the dynamical time on $\sim\SI{10}{kpc}$ scales is much shorter than the time resolution of most simulations. At low redshift, the dynamical time scale for galaxy clusters is of the same order as the SOAP time step\footnote{ For an $R\sim \SI{1}{Mpc}$, $M\sim 10^{14}\,\mathrm{M}_\odot$ cluster, the dynamical time scale $\delta t \sim \sqrt{R^3/GM} \sim \mathrm{10^9}\,\mathrm{yr}$. The corresponding redshift change at low redshift is $\delta z = - H(z)/a(z) \, \delta t \sim - H_0 \delta t \sim 0.1$.}. Whether interpolation would improve accuracy or introduce spurious structures for clusters is unclear. By not interpolating, our method preserves the local hydrodynamical structure at scales $\lesssim \SI{400}{Mpc}$\footnote{$\Delta z = 0.10$ corresponds to a comoving distance of $\SI{420}{Mpc}$.}. Moreover, interpolation requires identifying the main branch of cluster evolution between snapshots, which in turn requires accessing the merger tree history. This may introduce additional uncertainties and pathological behaviours \citep[see discussion in][]{smith_lightcone_2017}. The physical properties are only offset by $\Delta z < 0.05$, which is well within acceptable limits for our purposes.

\subsection{Sample selection and outlier removal}\label{sec:sample-selection}

In each lightcone, we selected clusters based on their X-ray flux $f_{X,\SI{0.1}{keV}\dash\SI{2.4}{keV}} > \SI{5e-12}{erg.s^{-1}.cm^{-2}}$ in the $0.1$--$2.4$ keV band (see Sect.~\ref{sec:sample-Lx} for details on the flux computation). A random Galactic plane was chosen for each lightcone, coordinates were transformed accordingly, and Galactic latitudes $|b|<20\degree$ were masked out to mimic the statistical behaviour of a Zone of Avoidance. To ensure no duplicated clusters, we then used the merger tree information to remove all duplicates\footnote{We do this by fetching all TopLeafIDs (the ID of the earliest progenitor in the main branch). If two clusters share the same TopLeafID, we retain only the one at lower redshift. Duplicates are very rare since no box repetition is used.}. After flux and latitude selection and duplicate removal, approximately $700$--$800$ clusters per lightcone remained, whereas M21 contains $\sim 350$. We note again that the reason for adopting the same selection process of M21 is to isolate its impact on the inferred dipole and does not reduce the predictive power of the simulation.

To understand why this discrepancy arises, we compared the distribution of projected X-ray concentration ($c_{\text{X}}$) between the selected simulated clusters and the sample used in M21. $c_{\text{X}}$ is defined as $L_\mathrm{core}/L_\mathrm{tot}$ where $L_\mathrm{core}$ and $L_\mathrm{tot}$ are the X-ray luminosities within $<0.15\ R_\mathrm{500c}$ and $<R_\mathrm{500c}$ respectively. We find that FLAMINGO clusters are less concentrated than the observed ones, where a significant portion of clusters with $c_{\text{X}}>0.5$ X-ray concentration is missing in FLAMINGO. The cause of this discrepancy is unclear, though it is diminished by accounting for projection effects. {The issue is potentially related to the fact that X-ray surveys tend to pick up the most concentrated cool-core clusters due to their high X-ray emission coming from their dense cores.} To create the most realistic catalogues and match statistics to M21, we choose to sort the clusters by X-ray concentration and consider only the first $\sim 300$--$400$ clusters for our analysis. The exact sample size depends on the relation and statistical methods used (Table~\ref{tab:scaling-relation-constants}). {This $c_\mathrm{X}$-based selection turns out to have no considerable effect on our final results as we show in Sect.~\ref{sec:effect-of-xray-concentration}.} (For further discussion on the $c_{\text{X}}$ discrepancy see Appendix~\ref{app:xray-concentration}.) The $c_{\text{X}}$ distribution and the differences therein are shown in Fig.~\ref{fig:xray-concentration}.

Strong scaling relation outliers can bias the obtained cosmological constraints, especially when small cluster samples are analysed in specific sky regions. Such outliers are absent in the M21 sample, as a sample cleaning was performed prior to the analysis (removing AGN-dominated clusters and highly disturbed mergers). To avoid this potential bias in our work, we consider the full-sky scaling relation fit and perform a $4\sigma_\mathrm{intr}$ clipping from the best-fit scaling relation (see Sect.~\ref{sec:method} for our power law fitting method) to remove outliers, where $\sigma_\mathrm{intr}$ is the fitted intrinsic scatter of each relation. For samples of $\lesssim 400$ clusters, it is unlikely that any object would scatter beyond the $4\sigma$ threshold by chance; thus, the removed clusters are likely to be peculiar cases not expected to follow universal scaling relations due to their physical characteristics. Outliers are removed for all three relations of interest, $L_\mathrm{X}\dash T$, $Y_\mathrm{SZ}\dash T$, and $M_\mathrm{gas}\dash T$. Typically, the number of outliers per lightcone is $<10$, i.e. $<3\%$ of the sample.

\subsection{Simulation samples and cluster properties} \label{sec:sample-property}

After lightcone creation and flux-based sample selection, a total of 1728 cluster catalogues were obtained, each containing $700$--$800$ clusters. Before evaluating their anisotropies, the robustness of the simulated samples must first be established. This section presents and compares our lightcone catalogues to the M20 sample and the FLAMINGO L1\_m9 runtime lightcones. The runtime lightcones provide a more accurate account of these quantities by yielding more precise redshifts and extracting properties from snapshots closer to the lightcone crossing redshift. (For more details on the comparison of the samples, see Appendix~\ref{app:sample-comparison} and the figures therein.)

\subsubsection{Redshift distribution and sky coordinates}
\label{sec:redshift-distribution-sky-coord}

\begin{figure}
    \centering
    \includegraphics[width=1\linewidth]{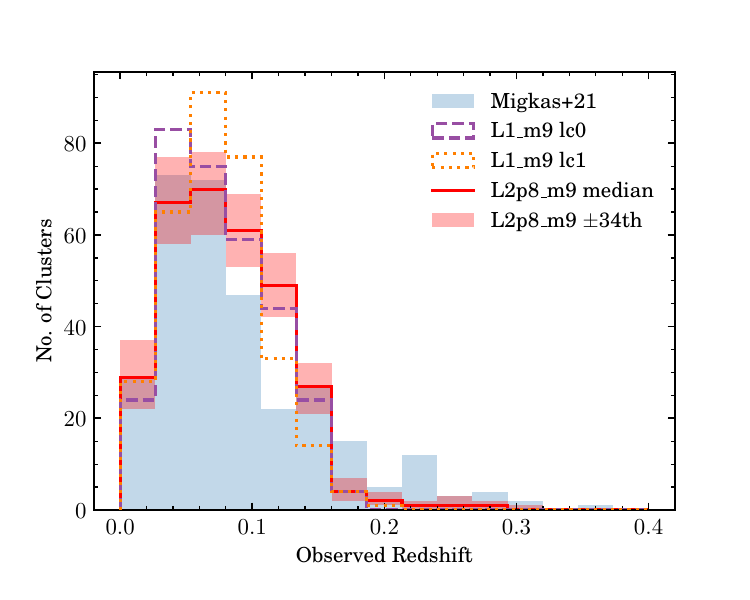}
    \caption{Average cluster redshift distribution of the 1728 
$(\SI{2.8}{Gpc})^3$ lightcones (red) compared with two runtime 
$(\SI{1.0}{Gpc})^3$ lightcones (purple, orange) and the M21 sample (blue). 
The most X-ray concentrated clusters from the simulation were selected to 
match the M21 sample size ($N = 313$).}
    \label{fig:redshift-distribution}
\end{figure}

Generally, the properties of the simulated lightcone samples are in good agreement with the M21 sample. The average redshift distribution of the 1728 simulated samples is shown in Fig.~\ref{fig:redshift-distribution} and compared with the M21 sample. The distribution extends up to $z \sim 0.30$ with a median of $z \sim 0.07$. Our simulated samples rarely extend above $z \sim 0.15$, while the M21 observed clusters exhibit a more pronounced tail at higher redshift. However, in both cases, clusters at $z \lesssim 0.15$ dominate the samples and thus primarily determine the probed scaling relation behaviour. 

\subsubsection{X-ray luminosity and flux}\label{sec:sample-Lx}

In SOAP, $L_\mathrm{X}$ is computed using the photoionisation code \textsc{Cloudy} \citep{ferland_2017_2017}, following the method of \citet{ploeckinger_radiative_2020}. The implementation is described in detail in \citet{braspenning_flamingo_2024}. The final cluster $L_\mathrm{X}$ is calculated as the sum of the X-ray luminosities of all particles within $R_\mathrm{500c}$. SOAP provides $L_\mathrm{X}$ values for three bands: $0.5$--$2$ keV, $0.2$--$2.3$ keV, and $2.3$--$8$ keV.

To select samples consistent with M20 and M21, which used fluxes in the $0.1$--$\SI{2.4}{keV}$ band, we started from the SOAP-provided $L_\mathrm{X}$ in the $0.2$--$\SI{2.3}{keV}$ band and converted it to the $0.1$--$\SI{2.4}{keV}$ band using an unabsorbed \texttt{apec} model in \texttt{XSPEC} \citep{arnaud_xspec_1999} and the cluster's temperature (see Sect.~\ref{sec:sample-T}). The observed flux was then computed from the converted $L_\mathrm{X}$, using $z_{\text{obs}}$ to account for the K-correction. This flux was used for sample selection. Although the flux-selected number of clusters remains higher than in M21, by this process we ensured that this discrepancy was not caused by incorrect flux estimation or the bandpass conversion.

However, for our analysis of the $\LT$ relation, we used the original SOAP $L_\mathrm{X}$ values in the $0.2$--$\SI{2.3}{keV}$ band. While not strictly identical to the $0.1$--$2.4$ keV band used in M21, the difference in $L_\mathrm{X}$ has a negligible impact on the angular variation of $\LT$, which is the focus of our study. Reprocessing the full pipeline using converted luminosities would introduce additional uncertainties and reduce reproducibility, without materially changing the anisotropy analysis. This choice avoids introducing
additional uncertainties from model-based conversions and ensures that our
results remain fully reproducible using the native SOAP outputs. The resulting $f_\mathrm{X}$ and $L_\mathrm{X}$ distributions of our simulated lightcone samples are in good agreement with both the L1\_m9 runtime lightcones and the observational data from M21 (See Appendix~\ref{app:cluster-properties-comparison} and figures therein). The typical\footnote{By “typical” we refer to the median of all distributions. The three reported numbers here and below are the medians of the minimum, maximum, and median values of the 1728 lightcones.} $L_\mathrm{X}$ distribution spans $L_\mathrm{X} = (6 \times 10^{41}$--$2 \times 10^{45})\,\SI{}{erg.s^{-1}}$, with a median of $\SI{e44}{erg.s^{-1}}$.

\subsubsection{Total integrated Compton $Y_{\text{SZ}}$ parameter}\label{sec:sample-YSZ}

The $Y_\mathrm{SZ}$ parameter was calculated by summing over the contribution of each gas particle inside the $5\times R_\mathrm{500c}$ aperture
\citep{schaye_flamingo_2023}:
\begin{align}
    Y_\mathrm{SZ}[\mathrm{kpc}^2] = \frac{\sigma_T}{m_e c^2} k_B\sum_i n_{e,i} T_{e,i}
    \frac{m_i}{\rho_i},
\end{align}
where $\sigma_T$ is the Thomson cross section, $c$ is the speed of light, $m_e$ is the electron mass, $k_B$ is the Boltzmann constant, $m_i$ and $\rho_i$ are the mass and density of particle $i$, and $n_{e,i}$ and $T_{e,i}$ are its electron number density and temperature. This numerical implementation corresponds to the $Y_\mathrm{SZ}$ definition used in M21, where the integral is performed along the line of sight and over the area.\footnote{In M21, $Y_\mathrm{SZ} = D_A^2 \int \dd\Omega \int_\mathrm{l.o.s.} \dd{l} \frac{\sigma_T}{m_e c^2} k_B n_e T_e$, where `l.o.s.' denotes the line of sight towards the cluster. The two definitions are equivalent. We neglect non-cluster line-of-sight contributions.} 

In M21, 260 clusters with $Y_\mathrm{SZ}$ S/N>2 were selected for analysis. For comparison, we used the same number and selected the 260 highest X-ray concentration clusters. Similar to $T$, the simulated $Y_\mathrm{SZ}$ distribution lies also on the lower end compared to the M21 data. The typical distribution of simulated lightcones spans $Y_\mathrm{SZ}=(0.2\dash 484.8)\,\mathrm{kpc^2}$, with a median value of $\SI{20.8}{kpc^2}$, whereas the M21 sample has $Y_\mathrm{SZ}=(1.3\dash 891.6)\,\mathrm{kpc^2}$ and the median is $\SI{33.7}{kpc^2}$.

\subsubsection{Gas mass}\label{sec:sample-Mgas}

The $M_\mathrm{gas}$ of a cluster is defined in SOAP as the sum of the masses of all gas particles within $R_{500c}$. The typical distribution of $M_\mathrm{gas}$ in the lightcones has a range of $M_\mathrm{gas} = (5.1\times10^{11} \dash 2.3\times 10^{14})\,\mathrm{M}_\odot$ with a median of $2.5\times10^{13}\,\mathrm{M}_\odot$. $M_\mathrm{gas}$ has not previously been used in any anisotropy analyses, but such work is forthcoming (Migkas et al., in prep.). In this study, the $\MT$ relation is employed to probe cosmic anisotropy for the first time, providing a reference point for future investigations. We included all gas particles in the computation of $M_\mathrm{gas}$ to avoid any bias, even though cold gas is invisible in X-ray observations. However, we have verified that the difference introduced by this choice is $\lesssim 1\%$.

\subsubsection{Temperature}\label{sec:sample-T}

The spectroscopic-like temperature was calculated following \citet{mazzotta_comparing_2004} as
\begin{align}
    T_\mathrm{sl} = \frac{\sum_i \rho_i m_i T_i^{1/4}}{\sum_i \rho_i m_i
    T_i^{-3/4}},\label{eq:spectroscopic-like temperature}
\end{align}
where $\rho_i$, $m_i$, and $T_i$ are the density, mass, and temperature of particle $i$. We use $T_\mathrm{sl}$ calculated within a core-excised aperture of $0.15\,R_\mathrm{500c} < r < R_\mathrm{500c}$. This was done to avoid the bias from cool-core clusters and to remain consistent with M21, which also excluded the cluster core using the $0.2$--$0.5\,R_{500}$ annulus. Only hot gas is visible in X-ray, and thus only particles with $T > 10^5\,\SI{}{K}$ are summed over.\footnote{In contract to $M_\mathrm{gas}$, this selection is essential, as including all gas particles leads to $T_\mathrm{sl}$ values that are lower by $2$--$3$ orders of magnitude, due to the strong weighting of cold, dense gas.}

It is well established that a systematic difference exists between the \textit{XMM-Newton} and \textit{Chandra} X-ray telescopes in measuring $T$ for the same clusters \citep{schellenberger_xmm-newton_2015, migkas_srgerosita_2024}. We find that the slope values of scaling relations using SOAP $T_\mathrm{sl}$ closely resemble results from previous observational studies that used \textit{XMM-Newton} temperatures. Thus, we consider $T_\mathrm{sl}$ to be equivalent to \textit{XMM-Newton} $T$. However, M21 used \textit{Chandra} $T$. To better match the scaling relation behaviour in M21, which is crucial for comparison, we converted $T_\mathrm{sl}$ to a \textit{Chandra}-like temperature using $T_\text{Cl} = T_\mathrm{sl}^{1/0.89}$, following \citet{schellenberger_xmm-newton_2015}. After this conversion, the M21 scaling relation behaviour is well replicated (see Sect.~\ref{sec:scaling-relation-behaviours}).

The above conversion was applied only when using the M21 statistical methodology (Sect.~\ref{sec:method-m21}), in order to enable direct comparison with the M21 anisotropy results. When the alternative statistical procedure (MCMC) was used to explore anisotropies in FLAMINGO (Sect.~\ref{sec:method-mcmc}), the default $T_\mathrm{sl}$ was retained.

The typical \textit{XMM-Newton}-like $T_\mathrm{sl}$ distribution in our lightcones spans $(0.6$--$10.7)\,\mathrm{keV}$ with a median of $\SI{3.1}{keV}$. M21's 313 clusters have a higher median temperature of $\SI{4.5}{keV}$. After converting $T_\mathrm{sl}$ to a \textit{Chandra}-equivalent value as discussed above, the temperature difference was significantly reduced. The resulting distribution spans $T_\text{Cl} = (0.57$--$14.34)\,\mathrm{keV}$, with a median of $\SI{3.58}{keV}$.

The remaining difference between the M21 and FLAMINGO temperatures could arise from several factors, including the different apertures used for temperature measurement, as well as systematic biases in temperature calibration. The fiducial selection of hot gas by $T>10^5\,\mathrm{K}$ might also contributes if the cut is set too high. Moreover, the under-representation of high-redshift clusters in our flux-limited, simulated samples, as previously discussed, also affects intrinsic cluster properties such as $T$, $L_\mathrm{X}$, $Y_\mathrm{SZ}$, $M_\mathrm{500c}$, and $M_\mathrm{gas}$, lowering their average values. We have verified that our high $c_\mathrm{X}$ selection (Sect.~\ref{sec:sample-selection}) does not cause the temperature offset, by examining the $T$ distribution of the lowest-$c_X$ clusters.

\subsubsection{Velocities}\label{sec:sample-velocities}

The three-dimensional velocity $\vb{v}$ of each cluster was taken to be the centre-of-mass velocity computed by SOAP, using all particles within $R_\mathrm{500c}$. The typical peculiar velocity spans $|\vb{v}| = (41$--$1436)\,\SI{}{km\,s^{-1}}$, with a median of $\SI{452}{km\,s^{-1}}$. The true bulk flow, in contrast to the values inferred from scaling relations, was estimated as the mean peculiar velocity of all clusters within a given redshift shell. The bulk flow profile as a function of redshift is presented in Sect.~\ref{sec:results-bf}.

\subsection{Matched scatter with M21}\label{sec:matched-scatter}

\begin{figure}
    \centering
    \includegraphics[width=\linewidth]{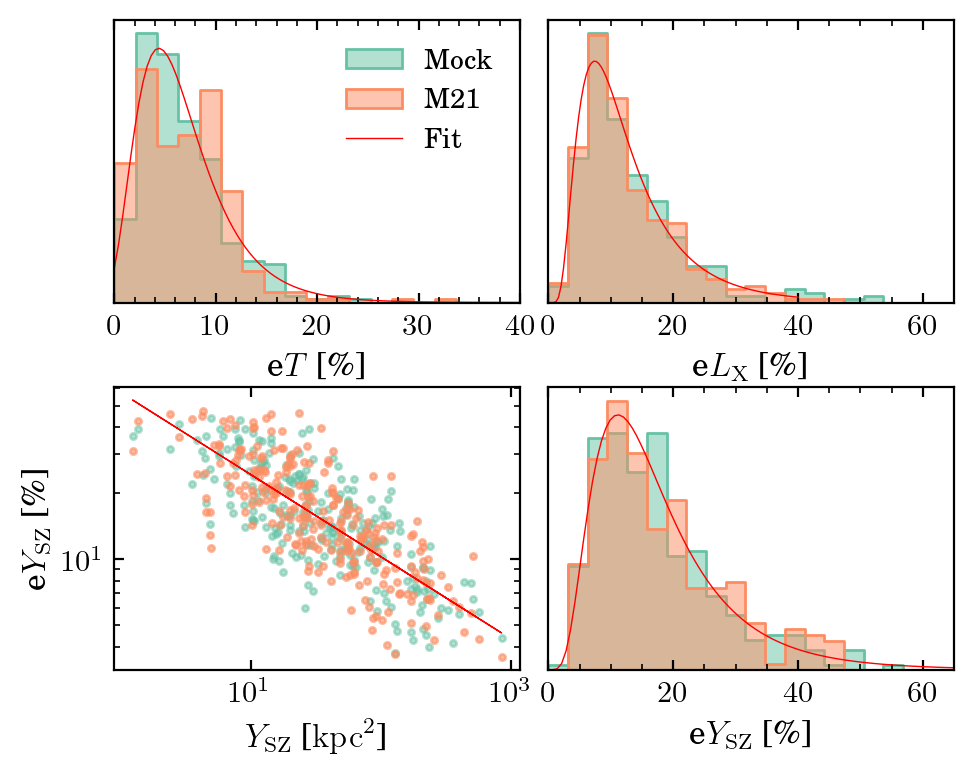}
    \caption{Mock instrumental uncertainties for one lightcone in $T$ (top left), $L_\mathrm{X}$ (top right), and $Y_\mathrm{SZ}$ (bottom right) compared with the M21 sample. The bottom-left panel shows the correlation between $Y_\mathrm{SZ}$ and its uncertainty $eY_\mathrm{SZ}$. In each panel, the mock uncertainty (green) represents one realisation drawn from the fitted distribution (red).} 
    \label{fig:instrumental-uncertainty}
\end{figure}

Our anisotropy analysis is highly sensitive to the scatter, as the inferred amplitude of a spurious cosmic anisotropy increases with larger scatter in the scaling relations (see Sect.~\ref{sec:discussion-cosmological-implication}). To minimise bias from a mismatch between the scatter in FLAMINGO and M21, we generated a complete set of separate lightcone catalogues with scaling relation scatter matched to that of M21.

Two sets of tests are presented in parallel throughout this work. The first is the default run, in which all values follow the original simulation prescriptions. The second includes added scatter, matched to the uncertainties in the M21 catalogue. This matched scatter test was applied only to the $\LT$ and $\YT$ relations and not to $\MT$; in other words, scatter was injected into $L_\mathrm{X}$, $Y_\mathrm{SZ}$, and $T$. The same uncertainty generation procedure is used for both $T_\mathrm{Chandra\text{-}like}$ and $T_\mathrm{sl}$.

Scatter was added using a two-step procedure. First, we amplified the deviation of the cluster property $Y$ ($L_\mathrm{X}$ and $Y_\mathrm{SZ}$) from the best-fit power law:
\begin{align}
    \log Y'_\mathrm{s} = [\log Y' - \log Y'_\mathrm{th}(X')]
    \frac{\sigma_\mathrm{intr, M21}}{\sigma_\mathrm{intr, FL}} + \log
    Y'_\mathrm{th}(X'), 
\end{align}
where $\log Y'_\mathrm{th}(X') = \log A_{YX} + B_{YX} \log X'$ denotes the logarithmic, normalised $Y$ from the fitted power law in terms of property $X$. The subscript $s$ marks quantities with injected scatter. Here, $\sigma_\mathrm{intr, FL}$ denotes the median intrinsic scatter across FLAMINGO lightcones (Table~\ref{tab:scaling-relation-behaviours}), while $\sigma_\mathrm{intr, M21}$ refers to the best-fit intrinsic scatter reported in M21. This amplification step matches the intrinsic scatter to the observed level while preserving the physical correlations introduced by the FLAMINGO baryonic physics model.

Second, we assigned a percentage-based instrumental uncertainty drawn from the M21 distribution. For $L_\mathrm{X}$ and $T$, where no strong correlation is seen between the uncertainty and the property. We fitted a log-normal distribution to the M21 uncertainties and sampled from it. For $Y_\mathrm{SZ}$, where a clear correlation is present, the dependence was explicitly modelled in the sampling. The measured value was then perturbed using the drawn uncertainty. The resulting uncertainty distributions and one example lightcone are shown in Fig.~\ref{fig:instrumental-uncertainty}. The log-normal fit accurately captures the observed behaviour. We note that the uncertainties were generated as percentages of the underlying property values. Thus the lower absolute values of $T_\mathrm{sl}$ do not lead to systematically higher relative uncertainties compared to $T_\text{Cl}$.

\section{Method}\label{sec:method}

We fitted a cluster scaling relation between $Y$ and $X$ with the following power law form:
\begin{align}\label{eq:scaling-relation}
    \frac{Y}{C_Y} E(z)^{\gamma_{YX}} = A_{YX} \qty(\frac{X}{C_X})^{B_{YX}},
\end{align}
where $C_X$ and $C_Y$ are predefined pivot points for the quantities $X$ and $Y$, $A_{YX}$ is the normalisation of the relation, and $B_{YX}$ is the slope. The constants $C_X$ and $C_Y$ were chosen to minimise the covariance between the fitted slope $B_{YX}$ and normalisation $A_{YX}$ in log space. Specifically, we set $C_X$ and $C_Y$ to the median values of $X$ and $Y$. The term $E(z) = \sqrt{\Omega_m(1+z)^3 + \Omega_\Lambda}$ is the evolution factor, that, together with the power index $\gamma_{YX}$, accounts for the redshift evolution of the scaling relation. The value of $\gamma_{YX}$ is fixed to the self-similar value \citep[e.g.][]{giodini_scaling_2013}. The precise value of $\gamma_{YX}$ is unimportant for our direction-dependent comparisons, given the low-$z$ nature of our sample and the nearly uniform redshift distribution across the sky. The constants used are summarised in Table~\ref{tab:scaling-relation-constants}. As in M21, we focus on relative angular variations and not absolute parameters, so selection effects can be safely ignored under the assumption of uniform sky coverage and complete detection above the flux threshold.

\begin{table}
\centering
\caption{Constants used for the scaling relation fitting.}
\label{tab:scaling-relation-constants}
\footnotesize
$
\begin{array}{cccccc}
\toprule
\text{Relation} & N & C_X & C_Y & \gamma & \text{Configuration}
\\ 
\midrule
\multirow{2}{*}{$\LT$} & 313 & 4.2 \, \text{keV} & 10^{44}\,\SI{}{erg.s^{-1}} & -1  & \text{M21}\\
    & 400 & 3.8 \, \text{keV} & 10^{44}\,\SI{}{erg.s^{-1}} & -1 & \text{MCMC}\\
\midrule
\multirow{2}{*}{$\YT$} & 260 & 3.7 \, \text{keV} & 20 \,
\text{kpc}^2 & 1  & \text{M21}\\
    & 400 & 3.2 \, \text{keV} & 20 \,
    \text{kpc}^2 & 1  & \text{MCMC}\\
\midrule
\multirow{2}{*}{$\MT$} & 300 & 4.2 \, \text{keV} & 3 \times 10^{13}
\, \mathrm{M}_\odot & 1 & \text{M21}\\
    & 400 & 3.6 \, \text{keV} & 3 \times
    10^{13} \, \mathrm{M}_\odot & 1 & \text{MCMC}\\
\bottomrule
\end{array}
$
\tablefoot{For each relation, we list the number of selected clusters $N$, the calibration constants $C_X$ and $C_Y$, and the fixed redshift evolution index $\gamma$. Two configurations are shown: one matching the M21 observational setup using \textit{Chandra}-like temperatures $T_\mathrm{Cl}$, and one using the original SOAP $T_\mathrm{sl}$ values for the MCMC analysis.}
\end{table}

\begin{table*}
\centering
\caption{Best-fit parameters of the scaling relations using the full lightcone sample.} 
\label{tab:scaling-relation-behaviours}
\footnotesize
$
\begin{array}{ccccccccc}
\toprule
\text{Relation ($Y\dash X$)} & A_{YX} & B_{YX} & \sigma_\mathrm{intr} &
A_{YX}^{(s)} & B_{YX}^{(s)} & \sigma_\mathrm{intr}^{(s)} &
\sigma_\mathrm{tot}^{(s)} & \text{Configuration}\\ 
\midrule
\addlinespace
\multirow{2}{*}{$\LT$} & 1.770^{+0.041}_{-0.032} & 2.220^{+0.070}_{-0.078} &
0.164^{+0.012}_{-0.009} & 1.762^{+0.049}_{-0.040} & 2.215^{+0.081}_{-0.084} &
0.240^{+0.018}_{-0.018} & 0.251^{+0.018}_{-0.016} & \text{M21}\\
\addlinespace
    & 1.754^{+0.016}_{-0.024} & 2.500^{+0.063}_{-0.069} &
    0.154^{+0.012}_{-0.004} & 1.738^{+0.032}_{-0.040} & 2.506^{+0.069}_{-0.087}
    & 0.228^{+0.012}_{-0.018} & 0.243^{+0.012}_{-0.015} & \text{MCMC}\\
\addlinespace
\midrule
\addlinespace
\multirow{2}{*}{$\YT$} & 1.094^{+0.015}_{-0.020} & 2.441^{+0.030}_{-0.033} &
0.110^{+0.006}_{-0.006} & 1.081^{+0.030}_{-0.025} & 2.448^{+0.051}_{-0.054} &
0.156^{+0.012}_{-0.012} & 0.189^{+0.010}_{-0.009} & \text{M21}\\
\addlinespace
    & 1.102^{+0.015}_{-0.015} & 2.748^{+0.032}_{-0.029} &
    0.108^{+0.004}_{-0.004} & 1.096^{+0.020}_{-0.020} & 2.763^{+0.048}_{-0.048} &
0.156^{+0.012}_{-0.006} & 0.195^{+0.009}_{-0.009} & \text{MCMC}\\
\addlinespace
\midrule
\addlinespace
\multirow{2}{*}{$\MT$} & 1.079^{+0.010}_{-0.010} & 1.827^{+0.039}_{-0.039} &
0.080^{+0.006}_{-0.006} & \dash & \dash & \dash & \dash & \text{M21}\\
\addlinespace
    & 1.086^{+0.010}_{-0.005} & 2.050^{+0.039}_{-0.039} &
    0.078^{+0.004}_{-0.008} & \dash & \dash & \dash & \dash & \text{MCMC}\\
\addlinespace
\bottomrule
\end{array}
$
\tablefoot{All fits are performed using the reduced $\chi^2$ regression method (Sect.~\ref{sec:method-m21-chi-square}). The reported uncertainties in $A$, $B$, and $\sigma_{\mathrm{intr}}$ correspond to the 16th and 84th percentiles across all lightcones. Superscript $^{(s)}$ denotes fits after injecting scatter to match the M21 sample (Sect.~\ref{sec:matched-scatter}). The `Configuration' column indicates the number of clusters and fitting constants used, as defined in Table~\ref{tab:scaling-relation-constants}.}
\end{table*}

\subsection{$H_0$ anisotropy: The M21 method}\label{sec:method-m21}

The first approach follows M21 and uses a $\chi^2$ minimisation procedure to fit $H_0$ in different sky patches. The variation in the best-fit scaling relation normalisation is then translated into an apparent $H_0$ variation.

\subsubsection{$\chi^2$ minimisation} \label{sec:method-m21-chi-square}

Consider a scaling relation $Y\dash X$ for a cluster sample, with each cluster labelled by an index $i$ and described by properties $Y_i$ and $X_i$. We fitted the parameters $A_{YX}$, $B_{YX}$, and the intrinsic scatter $\sigma^2_\mathrm{intr}$ in logarithmic space by minimising the following expression:
\begin{align}
    \chi_Y^2 = \sum_{i=1}^N \frac{\qty(\log Y_i'(z_{\mathrm{obs},i}) - \log
    A_{YX} - B_{YX}\log X_i')^2}{\sigma^2_{\mathrm{intr}} + 
    \sigma^2_{\log{Y},i} + B^2 \times \sigma^2_{\log{X},i}}, \label{eq:chi2}
\end{align}
where $z_{\mathrm{obs},i}$ is the observed redshift in the cosmic rest frame (Eq.~\ref{eq:zobs}), $Y_i' = E(z_{\mathrm{obs},i})^{\gamma_{YX}} Y_i / C_Y$, $X_i' = X_i / C_X$, and $N$ is the number of clusters. The terms $\sigma_{\log{Y},i}$ and $\sigma_{\log{X},i}$ denote the mock instrumental uncertainties in logarithmic space\footnote{We present results both with and without mock instrumental uncertainties. When uncertainties are omitted, $\sigma_{\log{Y},i} = \sigma_{\log{X},i} = 0$ in Eq.~\ref{eq:chi2} and \ref{eq:likelihood-function}, and the intrinsic scatter equals the total scatter. The total scatter is defined as $\sigma_\mathrm{tot} = \sqrt{\sigma^2_\mathrm{intr} + \sigma^2_{\log{Y},i} + B^2 \sigma^2_{\log{X},i}}$.}. Following \citet{maughan_lx-yx_2007}, the intrinsic scatter $\sigma^2_{\mathrm{intr},YX}$ is iteratively increased until the reduced chi-squared reaches unity, $\chi^2_\mathrm{res} \sim 1$\footnote{The reduced chi-squared is defined as $\chi^2_\mathrm{res} \equiv \chi^2_Y / (N - \mathrm{d.o.f.})$, with $\mathrm{d.o.f.} = 3$ corresponding to the free parameters $A_{YX}$, $B_{YX}$, and $\sigma_\mathrm{intr}$.}. We always take $Y$ to be the quantity with stronger cosmology dependence, and only squared deviations along the $Y$-axis are minimised.

\subsubsection{Sky scanning, uncertainties, and $H_0$ variation}\label{sec:method-scanning}

Cluster scaling relations were fitted within a cone of opening angle $\theta = 60\degree$ (for $\YT$) or $75\degree$ (for $\LT$ and $\MT$), and the resulting best-fit parameter was assigned to the cone centre $\vu{n}(l, b)$ to construct a sky map $p(\vu{n})$. The cone was scanned across the sky in steps of $4\degree$ in longitude and $2\degree$ in latitude. To enhance the contribution of clusters closer to the cone centre relative to those near its edge, each cluster was weighted by $\cos(\alpha_i)$, where $\alpha_i$ is its angular separation from the cone centre, following the method of M21. The cone size $\theta$ was chosen to balance the need for sufficient clusters in each patch and the desire for directional resolution. After the best-fit parameters are determined in all directions, uncertainties are estimated using bootstrap resampling with 500 realisations. The $1\sigma$ uncertainty $\sigma_p$ was taken as the $\pm34$th percentile around the best-fitting parameter $p$.

An anisotropy in $H_0$ affects all redshifts equally and does not alter the slope $B$ of the scaling relation. For relations between a cosmology-dependent quantity (such as $L_\mathrm{X}$, $Y_\mathrm{SZ}$, or $M_\mathrm{gas}$) and a cosmology-independent one (such as $T$), the variation in the normalisation $A(\vu{n})$ directly reflects directional variation in $H_0(\vu{n})$.

Taking $\LT$ as an example, the luminosity is defined by $L_\mathrm{X} = 4\pi D_L(z)^2 E(z)^{\gamma} f_X$, where the luminosity distance is $D_L(z) = (1 + z) \int_0^z c\,\dd{z'} / [H_0 E(z')]$. On the other hand, the scaling relation implies that for a given $T$, the expected value is $L_\mathrm{X} = A_{LT} T^{B_{LT}}$, which is cosmology-independent. Ignoring normalisation constants $C_L$ and $C_T$ for clarity, equating the two gives
\begin{align}
    L_\mathrm{X} = 4\pi D_L(z)^2 f_X E(z)^{\gamma} = A_{LT} T^{B_{LT}}.
\end{align}
Thus,
\begin{align}
    A_{LT} \propto D_L(z)^2 \propto H_0^{-2},
\end{align}
meaning any overestimation of $L_\mathrm{X}$ for a fixed $T$ (reflected in a higher $A_{LT}$) implies an underestimated $H_0$, meaning the true $H_0$ is higher.

The same argument applies to the $\YT$ and $\MT$ relations. Since $Y_\mathrm{SZ} \propto D_A^{2}(z)$ \citep[e.g.][]{giodini_scaling_2013} and $M_\mathrm{gas} \propto D_A^{5/2}(z)$ \citep[e.g.][]{sasaki_new_1996}, with $D_A(z) = (1 + z)^{-2} D_L(z)$, we obtain:
\begin{align}
    A_{YT} &\propto H_0^{-2}, \\
    A_{MT} &\propto H_0^{-5/2}.
\end{align}

The directional variation of $H_0$ can therefore be inferred from the normalisation map of any of the three scaling relations:
\begin{align}
    \frac{H'_0(\vu{n})}{H_0} = \qty(\frac{A_{YX}(\vu{n})}{A_{YX,\mathrm{all}}})^{\alpha},
    \label{eq:h0-MT}
\end{align}
where $\alpha = 1/2$, $1/2$, and $2/5$ for the $\LT$, $\YT$, and $\MT$ relations, respectively. Here, $H_0'(\vu{n})$ denotes the direction-dependent Hubble constant, $H_0$ the isotropic value, $A_{YX}(\vu{n})$ the local normalisation, and $A_{YX,\mathrm{all}}$ the best-fit normalisation from all clusters in the same lightcone. We emphasise that we do not constrain the absolute value of $H_0$ from cluster scaling relations, but only its relative variation across the sky.

\subsubsection{Dipole anisotropy statistical significance}\label{sec:dipole-significance}

To assess the statistical significance of a dipole anisotropy, we analysed the sky map of the scaling relation normalisation $A(\vu{n})$, along with its bootstrapped uncertainty $\sigma_A(\vu{n})$. The significance of a dipole pointing in a given direction $\vu{n}$ was quantified as the difference in $A(\vu{n})$ and its value in the antipodal direction $-\vu{n}$, in units of the combined uncertainty:
\begin{align}\label{eq:chi2-statistical-significance}
    n_\sigma(\vu{n}) = \frac{A(\vu{n}) - A(-\vu{n})}
    {\sqrt{\sigma_A^2(\vu{n}) + \sigma_A^2(-\vu{n})}}.
\end{align}
By definition, $n_\sigma(\vu{n}) = -n_\sigma(-\vu{n})$. A negative value indicates that $A$ is lower in the direction $\vu{n}$ than in its opposite region. The amplitude of the $H_0$ dipole for each lightcone is then defined as the maximum directional difference in $H_0' / H_0$, i.e.
\begin{align}
\Delta H_0[\%] = [H_0'(\vu{n}) - H_0'(-\vu{n})] / H_0, \label{eq:DH0}
\end{align}
where $\vu{n} = \mathrm{arg\,max}\,n_\sigma(\vu{n})$ is the direction of maximal significance. Our approach differs slightly from that of M21, which compared each region to the rest of the sky rather than to its antipode. By focusing on direct dipole-level comparisons, our method is less prone to underestimating the amplitude of $H_0$ variation.

The statistical significance of the observed dipole was not taken from the maximum value of $n_\sigma(\vu{n})$ directly, as this would be biased by the look elsewhere effect (selecting the most extreme fluctuation across many directions). Instead, the significance is quantified by comparing against the FLAMINGO lightcones: we compute the fraction of them that yield a higher maximum $n_\sigma(\vu{n})$ (and corresponding $\Delta H_0$) than the observed value. The statistical method for this comparison is described in Sect.~\ref{sec:method-probability}. This defines a robust $p$-value that captures the probability of such an anisotropy arising in a statistically isotropic universe. The procedure mirrors that of M21, where mock catalogues were used to evaluate the significance of the detected dipole.

\subsection{$H_0$ anisotropy: MCMC full likelihood analysis}\label{sec:method-mcmc}

The second approach uses a full-sky MCMC fit in which $H_0$ is allowed to vary within the parametrisation of each scaling relation. This serves as a cross-validation of the $\chi^2$ sky-patch method (the M21 method) described above. A dipole-like $H_0$ variation is modelled as $H_0'/H_0 = 1 + \delta \cos \alpha_i(\phi_H, \lambda_H)$, where $\delta$ is the relative amplitude of the anisotropy and $\alpha_i$ is the angular separation between the $i$-th cluster and the dipole direction $(\phi_H, \lambda_H)$, given in simulated Galactic coordinates. A total of six parameters were constrained in the fit: $\delta$, $\phi_H$, $\lambda_H$, $A_{YX}$, $B_{YX}$, and $\sigma_\mathrm{intr}$. We sampled the likelihood function
\begin{align}\label{eq:likelihood-function}
    \ln\mathcal{L}_\mathrm{YX} = -\frac{1}{2} \chi^2_Y + \sum_i^N \ln(\sigma_{\mathrm{intr}}^2 + 
    \sigma^2_{\log{Y},i} + B^2\times \sigma^2_{\log{X},i}),
\end{align}
where $\chi^2_Y$ is defined in Eq.~\ref{eq:chi2}, and $N$ is the number of
clusters in the fit.

For each parameter, we report the posterior median, with the 16th and 84th percentiles defining the uncertainty interval. The amplitude of the anisotropy is $\Delta H_0 = 2\delta$, to be consistent with the M21 method (Eq.~\ref{eq:DH0}). The statistical significance $n_\sigma$ of the anisotropy was estimated by dividing the median $\delta$ by its lower uncertainty. This reflects the number of standard deviations separating the result from isotropy ($\delta = 0$). As with the M21 method, the per-lightcone $n_\sigma$ values were not used to infer significance directly, due to the look elsewhere effect (Sect.~\ref{sec:dipole-significance}). Only the distribution of anisotropy amplitudes across all lightcones is compared with the M21 result to determine statistical relevance. 

We used the \textsc{Emcee} package developed by \citet{foreman-mackey_emcee_2013} for MCMC sampling. The integrated autocorrelation time $\tau$ \citep{goodman_ensemble_2010} was monitored to ensure convergence, and all chains were run to lengths exceeding $50\tau$.

\subsection{Modelling bulk flows}\label{sec:method-bulk-flow}

Peculiar velocities influence cluster scaling relations by altering the observed redshift $z_\mathrm{obs}$ away from the cosmological redshift $z_\mathrm{cos}$. While random motions add to the intrinsic scatter without introducing directional effects, a coherent bulk flow can mimic an $H_0$ anisotropy in the scaling relations \citep{migkas_cosmological_2021, migkas_galaxy_2025}. This degeneracy is strongest at low redshifts, where peculiar velocities are comparable in magnitude to the Hubble expansion. To isolate the effect of bulk flows, we assume an isotropic $H_0$ throughout the bulk flow analysis.

The bulk flow was parametrised by its amplitude, $u_\mathrm{bf}$, and direction, $(l, b) = (\phi_\mathrm{bf}, \lambda_\mathrm{bf})$. For a given bulk flow, the observed redshift, $z_\mathrm{obs}$, was corrected using \citep{dai_measuring_2011}
\begin{align}
    z_\mathrm{cor} = \frac{z_\mathrm{obs} - z_\mathrm{bf}}{1 + z_\mathrm{bf}}\label{eq:bf-modeling},
\end{align}
where $z_\mathrm{cor}$ approximates the redshift in the absence of bulk motion. This equation follows from inverting Eq.~\ref{eq:zobs} for $z_\mathrm{cos}$. The bulk flow redshift shift, $z_\mathrm{bf}$, was calculated from the non-relativistic Doppler formula, $z_\mathrm{bf} = (u_\mathrm{bf} \cos \alpha) / c$, where $\alpha$ is the angular separation between the bulk flow direction $(\phi_\mathrm{bf}, \lambda_\mathrm{bf})$ and the cluster sky position $(\phi, \lambda)$. For bulk flow amplitudes $u_\mathrm{bf} < \SI{e3}{km\,s^{-1}}$, the accuracy\footnote{Eq.~\ref{eq:bf-modeling} is not exact, as the cluster peculiar velocity includes both a coherent bulk flow and a residual component. Thus, $v_\mathrm{pec} \neq u_\mathrm{bf}$ and $z_\mathrm{pec} \neq z_\mathrm{bf}$, but the approximation remains accurate at the $10^{-5}$ level.} of this non-relativistic treatment is sufficient.

We constrained the bulk flow parameters using MCMC to maximise the full likelihood function defined in Eq.~\ref{eq:likelihood-function}. At each sampling step, $z_\mathrm{cor}$ was recalculated using Eq.~\ref{eq:bf-modeling}, and the derived distances were used to modify the cosmology-dependent quantities: $L_\mathrm{X}[D_L(z_\mathrm{cor})]$, $Y_\mathrm{SZ}[D_A(z_\mathrm{cor})]$, and $M_\mathrm{gas}[D_A(z_\mathrm{cor})]$. The posterior median is quoted for each parameter, with the 16th and 84th percentiles giving the lower and upper uncertainty bounds.

\subsection{Probability estimation}\label{sec:method-probability}

As mentioned in Sect.~\ref{sec:dipole-significance}, the $n_\sigma$ or $\Delta H_0$ of a single lightcone (realisation) cannot be directly interpreted as the significance of a cosmic anisotropy due to the look-elsewhere effect. To assess whether the M21 result is in tension with $\Lambda$CDM, we instead consider the joint distribution of $\Delta H_0$ and $n_\sigma$ across all lightcones.

To estimate the probability of a certain outcome in the 2D parameter space ($\Delta H_0, n_\sigma$), we used two complementary approaches: kernel density estimation (KDE) and an extreme value statistics (EVS)-based method \citep{davison_models_1990, behrens_bayesian_2004,hu_bayesian_2024}. KDE is more accurate when the target point is not extremely rare (e.g. $p \gtrsim 1/N$), whilst EVS is suitable to estimate much lower probability and can extrapolate to rare events. In EVS, we applied the Peak-Over-Threshold approach, relying on the theorem of \citet{pickands_statistical_1975}, along with \citet{balkema_residual_1974}, which asserts that when the threshold is set high enough, the likelihood of surpassing it is governed by a generalized Pareto distribution (GPD). In our case of 1728 lightcones, a probability $\lesssim 1/1728 = 6 \times 10^{-4}$ will be better estimated by the EVS-POT method (hereafter EVS).
 
We applied KDE directly to the 2D distribution of $\Delta H_0$ and $n_\sigma$ of the detected anisotropies of all lightcones to estimate the rarity of the M21 findings within FLAMINGO.\footnote{M21 used only $n_{\sigma}$ to determine the $p-$value of the observed anisotropy, which can potentially underestimate the statistical significance of their findings.} The probability for a target point, $P_0$, was calculated by integrating the probability mass above regions where the KDE value exceeds that at the target:
\begin{align}
P_{\text{KDE},0} = 1 - \iint_{f > f_{0}} f(\Delta H_0, n_\sigma) \, d\Delta H_0
\, dn_\sigma.
\end{align}
Grid resolutions were tested to ensure numerical convergence. While effective near the main probability mass, KDE becomes unreliable in the sparsely populated tails, which is often our case, where the result depends heavily on grid resolutions and local distribution.

To address the limitations of KDE in the tails, we projected the 2D data onto 1D with some function $x=f(\Delta H_0, n_\sigma)$ and use the EVS approach to model the extreme values of $x$. We use a fiducial dimension reduction function, the lower uncertainty bound $x = \Delta H_0 - \Delta H_0 / n_\sigma$. The GPD is fitted to exceedances $(x-u)$ above a high threshold $u$, and scaled to match the survival probability at the threshold $P(x=u)$. This ensures a seamless transition between the bulk and EVS in the tail. The probability of target point $x_0$ is then given by 
\begin{align}
    P_{\text{EVS},0} = \left\{ \begin{aligned}
        &P(x>x_0), &x_0\leq u,
        \\
        &P(x>u)P_\text{GPD}(x>x_0|x>u), &x_0>u,
    \end{aligned} \right.\label{eq:projected-evs}
\end{align}
where $P_\text{GPD}(x>x_0|x>u)$ is normalised as $P_\text{GPD}(x>x_0|x>x_0)=1$, and $P(x>x_0)$ is the number of samples with $x>x_0$ divided by the total number of samples. The threshold is selected to give a good fit in the tail, usually the 95th percentile of $x$. The resulting estimation might depend on the choice of threshold, therefore we always vary the percentile to ensure convergence. Alternative dimension reduction functions were tested (see Appendix~\ref{app:about-evs}) without a significant change in the conclusions. The projection and EVS combination was also applied for bulk flow analysis to assess the discrepancy between FLAMINGO and M21 bulk flows (see Sect.\ref{sec:results-bf} for more details).

By combining KDE with the projected EVS method, we achieved a robust probability estimates across the entire range. This allows for a quantitative assessment of the rarity of the M21 anisotropy in FLAMINGO lightcones.

\begin{figure*}
    \centering
    \begin{subfigure}[b]{0.48\linewidth}
        \centering
        \includegraphics[width=\linewidth]{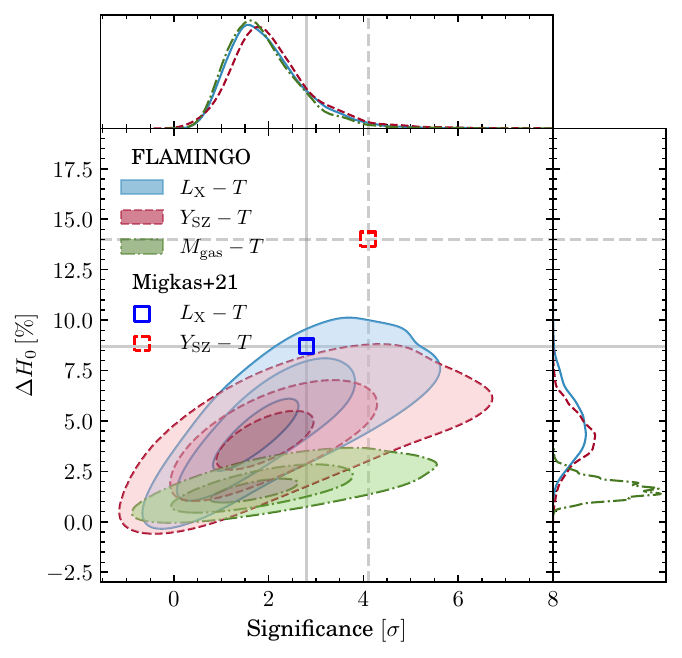}
        \subcaption{M21 method, no scatter}
        \label{fig:h0-result-chi2} 
    \end{subfigure}
    \begin{subfigure}[b]{0.465\linewidth} 
        \centering
        \includegraphics[width=\linewidth]{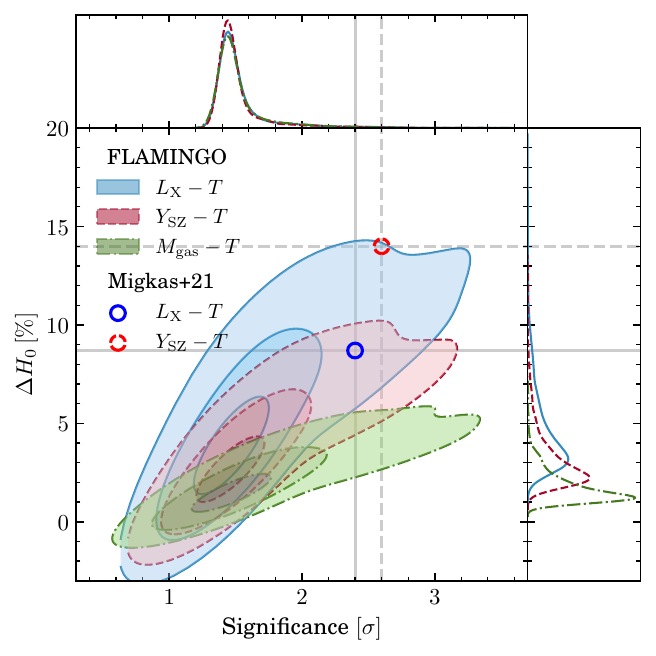}
        \subcaption{MCMC method, no scatter}
        \label{fig:h0-result-MCMC} 
    \end{subfigure}
    \begin{subfigure}[b]{0.48\linewidth}
        \centering
        \includegraphics[width=\linewidth]{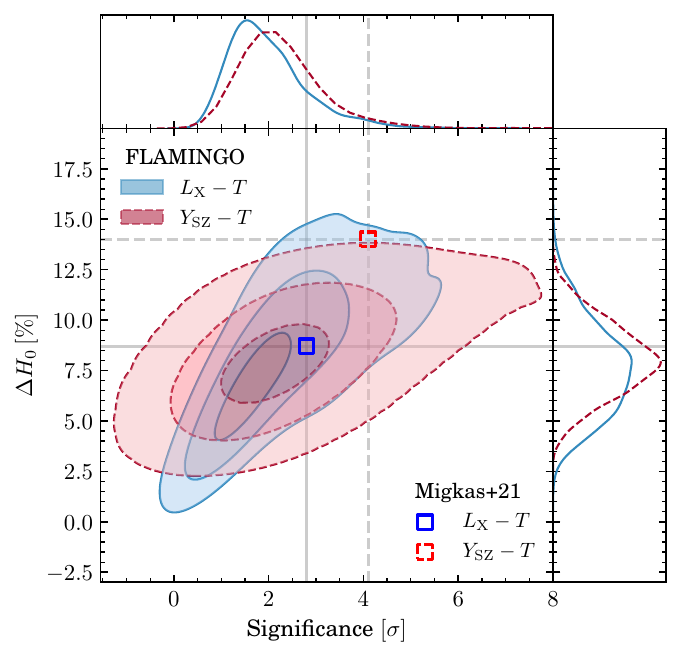}
        \subcaption{M21 method, matched scatter}
        \label{fig:h0-result-chi2-scatter} 
    \end{subfigure}
    \begin{subfigure}[b]{0.465\linewidth}
        \centering
        \includegraphics[width=\linewidth]{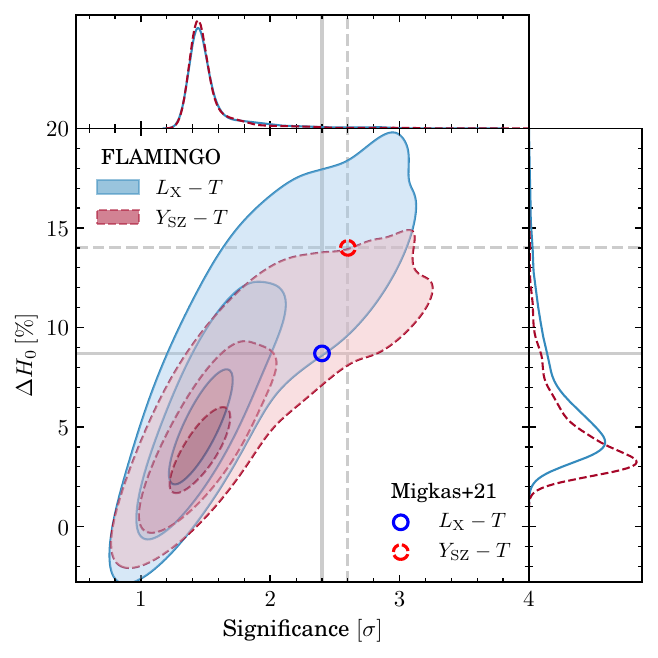}
        \subcaption{MCMC method, matched scatter}
        \label{fig:h0-result-MCMC-scatter} 
    \end{subfigure}
    \caption{Distribution of $H_0$ variation and its statistical significance from 1728 lightcones using the M21 scanning method (left) and the MCMC method (right) with (bottom) and without (top) mock scatter. The blue solid, red dashed, and green dash-dotted contours correspond to constraints from the $\LT$, $\YT$, and $\MT$ relations, respectively. Contour levels indicate 39\%, 86\%, and 98.9\% probability regions (equivalent to 1, 2, and 3$\sigma$ in two dimensions). Blue (red) squares mark the M21 results from $\LT$ ($\YT$), including their measured statistical significance. Blue (red) circles show the M21 results obtained from isotropic Monte Carlo realisations for $\LT$ ($\YT$). {While the $\LT$ results fall within the $3\sigma$ contours in every configuration, the $\YT$ results fall well outside the contours for the no-scatter case, and roughly at the $3\sigma$ limit for the matched scatter case.}}
    \label{fig:h0-result}
\end{figure*}

\section{Results}\label{sec:results}

\subsection{General behaviour of scaling relations}\label{sec:scaling-relation-behaviours}

Realistic scaling relations are a crucial aspect of our lightcone samples, they form the basis for evaluating anisotropy. The fitted parameters are summarised in Table~\ref{tab:scaling-relation-behaviours}. In this section, we compare our results with observations reported in the literature. We note that following the approach in M21, the default fitting parameters presented use the converted $T_\text{Cl}$ instead of $T_\mathrm{sl}$ for all three relations (Sect.~\ref{sec:sample-T}). 

The $\LT$ relation gives a slope of $B_{LT}=2.220^{+0.070}_{-0.078}$, steeper than the simplest self-similar model ($B_\mathrm{LT}=1.5$ for the soft band $L_{\text{X}}$) and fully consistent with M20 ($2.102\pm0.064$) and M21 ($2.086^{+0.073}_{-0.065}$). The slope is slightly shallower than studies using \textit{XMM-Newton} temperatures (\citet[e.g.][$2.73\pm0.13$]{chen_statistics_2007}, \citet[][$2.52\pm0.10$]{eckmiller_testing_2011}, \citet[][$3.110\pm0.422$]{lovisari_x-ray_2020}). Using the original SOAP $T_{\text{sl}}$ yields $B_{LT} = 2.500^{+0.063}_{-0.069}$ that better matches the higher values obtained in the above studies. The matched scatter test returns a higher scatter of $\sigma_\mathrm{intr}^{(s)} = 0.240^{+0.018}_{-0.018}$ ($\sigma_\mathrm{intr}^{(s)} = 0.228^{+0.012}_{-0.018}$ for $T_\mathrm{sl}$), reproducing the M21 scatter ($0.233 \pm 0.016$) {as well as independent observations (e.g. \citet[][0.28]{pratt_galaxy_2009}, \citet[][0.27]{eckmiller_testing_2011}, \citet[][0.26]{lovisari_scaling_2015})} while keeping the normalisation and slope unchanged. 

The $\YT$ relation gives the slope $B_{YT}=2.441^{+0.030}_{-0.033}$, consistent with M21 ($2.546^{+0.071}_{-0.067}$) and other \textit{Chandra} temperature observations (e.g. \citet[][$2.64\pm0.28$]{morandi_x-ray_2007}). The $\YT$ slope of our lightcones is very close to the self-similar model prediction $B_\mathrm{YT}=2.5$. Using the original $T_{\text{sl}}$, the simulated $\YT$ relation gives $B_{YT} = 2.748^{+0.032}_{-0.029}$ which is fully compatible with \citet[][$2.82\pm0.18$]{planck_collaboration_planck_2011}. Similar to $\LT$, the simulated lightcones give $25\%$ less scatter than observed in M21 ($0.146\pm 0.013$), with an intrinsic scatter of $\sigma_\mathrm{intr} = 0.110^{+0.006}_{-0.006}$. With matched scatter, we obtain $\sigma_\mathrm{intr}^{(s)} = 0.156^{+0.012}_{-0.012}$ for $T_\mathrm{Cl}$ and $0.156^{+0.012}_{-0.006}$ for $T_\mathrm{sl}${ which matches the intrinsic scatter of M21 and other observations (e.g. \citet[][$0.139$]{morandi_x-ray_2007}, \citet[][$0.14\pm0.02$]{planck_collaboration_planck_2011}).} The normalisation and slope remain the same. 

For the $\MT$ relation we obtain a slope of $B_{MT}=1.827\pm 0.039$, steeper than the self-similar model $B_{MT}=1.5$. Using original SOAP $T_\mathrm{sl}$, we find $B_{MT} = 2.050^{+0.039}_{-0.039}$, fully consistent with past \textit{XMM-Newton} measurements (e.g. \citet[][$2.10\pm0.11$]{arnaud_calibration_2007}, \citet[][$1.986\pm0.111$]{croston_galaxy-cluster_2008}, \citet[][$1.859\pm0.187$]{zhang_locuss_2008}). The intrinsic scatter for $\MT$ in our lightcones is $\sigma_\mathrm{intr} = 0.080^{+0.006}_{-0.006}$, making it the tightest relation of the three. As M21 does not measure $M_\mathrm{gas}$, no matched scatter test is applied here.

As shown above, the slopes of our scaling relations are consistent with those of M21. This ensures that both the simulated and observed relations respond similarly to bulk flows or statistical noise. The matched scatter tests preserve the original slopes and normalisations, while successfully reproducing the intrinsic and total scatter levels reported in M21. This correction is essential for a direct and unbiased assessment of anisotropy amplitudes.

\subsection{$H_0$ anisotropy} \label{sec:results-h0-m21}

The anisotropy constraints from the FLAMINGO lightcones, in comparison with the M21 results, are shown in Fig.~\ref{fig:h0-result} and \ref{fig:h0-result-joint-MCMC}. Table~\ref{tab:h0-$p$-value-summary} lists the $p$-values of the tension between the M21 detection and the FLAMINGO lightcones. The joint EVS distributions combining $\LT$ and $\YT$ are shown in Fig.~\ref{fig:evs-all-joint}, with per-relation analyses provided in Appendix~\ref{app:about-evs}.

\subsubsection{Constraints using $\LT$, $\YT$, or $\MT$}\label{sec:results-h0-per-relation}

For the $\LT$ relation, the M21 method yields a moderate $\Delta H_0$ peak at $4.3\%$ and a median dipole significance near $1.7\sigma$ (Fig.~\ref{fig:h0-result-chi2}). The observed M21 square lies above the main simulated region, with $p = 0.021$ (KDE) and $p = 0.025$ ($2.25\sigma$, EVS), suggesting a mild tension. However, once realistic scatter is added (Fig.~\ref{fig:h0-result-chi2-scatter}), the distribution shifts upward to $\sim 8\%$, overlapping the M21 point. This is reflected in the $p$-values, which rise to $p = 0.33$ (KDE) and $p = 0.23$ ($1.20\sigma$), indicating that the $\LT$ dipole is consistent with $\Lambda$CDM under observational scatter in the $\LT$ relation. 

The MCMC analyses of $\LT$ in our lightcones produce a more isotropic distribution (Fig.~\ref{fig:h0-result-MCMC}), peaking at $\Delta H_0 \sim 3.0\%$ and $n_\sigma \sim 1.4$. The M21 circle lies moderately above this region, with $p = 0.055$ (KDE) and $p = 0.032$ ($2.14\sigma$, EVS). After adding scatter (Fig.~\ref{fig:h0-result-MCMC-scatter}), the lightcone distribution shifts to $\sim 4\%$, but the $p$-values remain low: $p < 5.8 \times 10^{-4}$ (KDE) and $p = 0.085$ ($1.72\sigma$, EVS). In this case, scatter introduces more high $\Delta H_0$ tails, but the M21 result stays just outside the bulk. The mild increase in tension compared to the M21 method may reflect the broader likelihood sensitivity of the MCMC approach (see Sect.~\ref{subsec:method-compare} for a comparison between the two methods). The agreement with isotropy in $\LT$ is expected, as M21 also found only a mild
tension ($2.4\sigma$) in this relation.

The $\YT$ relation, by contrast, exhibits a stronger anisotropic signal in all cases. Without added scatter, the M21 square lies far outside the simulated region, with the KDE estimation bounded only from above, $p < 5.8 \times 10^{-4}$ (KDE)\footnote{The M21 point lies too far from the main probability mass, making KDE unreliable. We can place only an upper bound of $p < 5.8 \times 10^{-4}$ corresponding to the fraction $1/1728$, as none of the FLAMINGO lightcones exhibits a higher anisotropy.} and $p = 1.8 \times 10^{-10}$ ($6.38\sigma$, EVS), both are extremely low. The lightcone distribution itself peaks at $\Delta H_0 \sim 4.0\%$, significantly below the M21 value of $14\%$, as shown in Fig.~\ref{fig:h0-result-chi2}. With added scatter, the lightcones shift up to $\sim 8\%$ (Fig.~\ref{fig:h0-result-chi2-scatter}), and the tension is notably reduced: $p = 0.0013$ ($3.22\sigma$) in EVS and $p < 5.8\times 10^{-4}$ with KDE. The KDE result remains disproportionately low due to local sparsity in that region, despite the smoothed contours in Fig.~\ref{fig:h0-result-chi2-scatter} visually resembling a $3\sigma$ level. This highlights a limitation of KDE. Its sensitivity to the density of neighbouring points makes it less reliable when evaluating tail probabilities. In this case, the projected EVS method provides the more robust interpretation. 

The MCMC results for $\YT$ are similar in trend. Without added scatter, the lightcones peak at $\Delta H_0 \sim 2.2\%$ (Fig.~\ref{fig:h0-result-MCMC}), and the M21 circle is well beyond the contours: $p = 2.8 \times 10^{-4}$ ($3.64\sigma$). After scatter matching (Fig.~\ref{fig:h0-result-MCMC-scatter}), the peak shifts to $\sim 3\%$, and the M21 circle sits closer to the lightcone bulk. The $p$-values reduce to $p = 0.0049$ ($2.82\sigma$, EVS) and $p = 0.0014$ (KDE). The rarity slightly exceeds the isotropic mock sample analyses in M21 which reported $2.6\sigma$.

The $\MT$ relation, included for the first time in such anisotropy analysis, consistently shows the lowest $\Delta H_0$ values and the narrowest distributions on the isotropic lightcones. In the M21 method (Fig.~\ref{fig:h0-result-chi2}), $\Delta H_0$ peaks at $1.5\%$, and the dipole significance remains below $2\sigma$. Similarly, in the MCMC method (Fig.~\ref{fig:h0-result-MCMC}), $\Delta H_0$ peaks at $1.2\%$ and $n_\sigma \sim 1.4$, with a visibly tighter contour. These results identify $\MT$ as the most stable tracer of isotropy among the three relations. Since the relation was not presented in M21, we do not add scatter or compute $p$-values, but its low scatter and minimal anisotropy make it a promising probe for future studies, especially given the anisotropy amplitude being sensitive to scatter.

Altogether, these results show that scatter in cluster scaling relations significantly contributes to the fitted apparent anisotropy. For $\LT$, scatter could largely accounts for the discrepancy. For $\YT$, while scatter mitigates much of the tension, the remaining signal still places the M21 measurement in the upper tails of the lightcone distribution. $\MT$ offers a promising future prospect as a more sensitive anisotropy probe. The overall trends are consistent across methods (M21 or MCMC) and probability estimators (KDE or EVS), supporting the conclusion that scatter matters and that tension with $\Lambda$CDM remains in the $\YT$ detection.

\subsubsection{Joint analysis of both $\LT$ and $\YT$}\label{sec:results-h0-joint-analysis}

As different scaling relations may be affected by distinct systematics but all reflect the same underlying cosmology, a joint analysis can provide a more robust detection of a true cosmological anisotropy. However, past studies have shown a positive correlation between the scatter in $L_\mathrm{X}$ and $Y_\mathrm{SZ}$ at fixed mass \citep{nagarajan_weak-lensing_2019} or fixed temperature {(M21)}, likely due to the interplay of physical processes inside of clusters. As a result, if a cluster is upscattered in $\LT$, it is also likely to be upscattered in $\YT$. This correlation can cause random scatter to align and thus resemble a cosmological signal. M21 accounted for this by injecting the expected scatter correlation into their simulated samples used for significance estimation. When this correlation was included, the statistical significance of the dipole was found to be $5.9\sigma$. Ignoring the correlation led to an underestimated significance of $5.4\sigma$.

In this work, the correlation is built-in with the cluster physics in FLAMINGO. Therefore, we straightforwardly perform a joint likelihood MCMC analysis by sampling the combined log-likelihood function, 
\begin{align}
\log \mathcal{L} = \log \mathcal{L}_\mathrm{LT} + \log \mathcal{L}_\mathrm{YT},
\end{align}
where $\log \mathcal{L}_\mathrm{LT}$ and $\log \mathcal{L}_\mathrm{YT}$ are given by Eq.~\ref{eq:likelihood-function}. This result can be directly compared to the final constraints reported in M21 that are obtained by similarly multiplying the likelihoods of $L_{\mathrm{X}}$ and $Y_{\mathrm{SZ}}$. In our samples and M21, the anisotropy signals are boosted by correlated scatter in the same manner. Therefore, the effect is cancelled to allow for a probability assessment unbiased to this effect.

\begin{figure}
    \centering
    \includegraphics[width=1\linewidth]{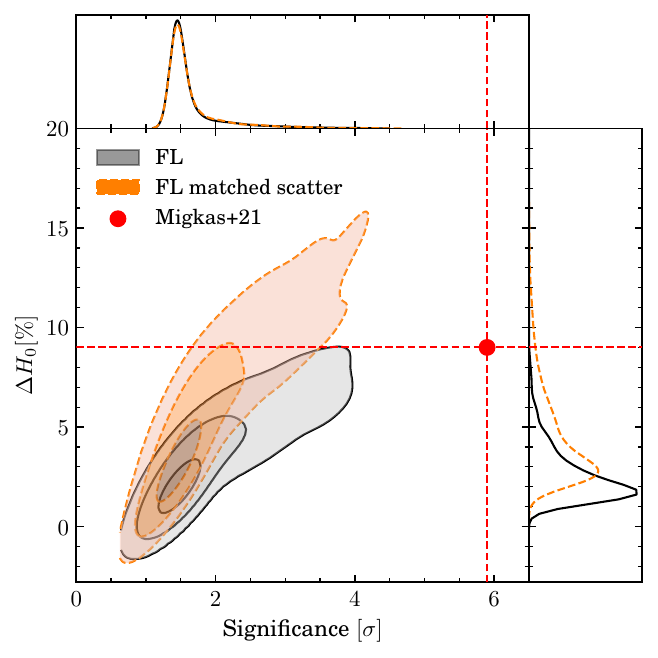}
    \caption{Joint MCMC constraint from the $\LT$ and $\YT$ relations compared to the combined M21 result (red dot). The orange contour includes injected scatter, while the grey does not. Contours show the 39\%, 86\%, and 98.9\% confidence regions.}
    \label{fig:h0-result-joint-MCMC}
\end{figure}

\begin{figure*}
    \centering
    \begin{subfigure}[b]{0.48\linewidth}
        \centering
        \includegraphics[width=\linewidth]{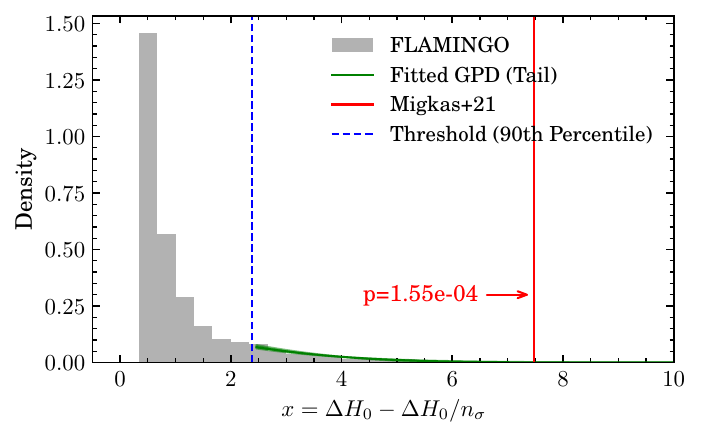}
        \subcaption{No scatter, MCMC}
        \label{fig:evs-joint}
    \end{subfigure}
    \begin{subfigure}[b]{0.48\linewidth}
        \centering
        \includegraphics[width=\linewidth]{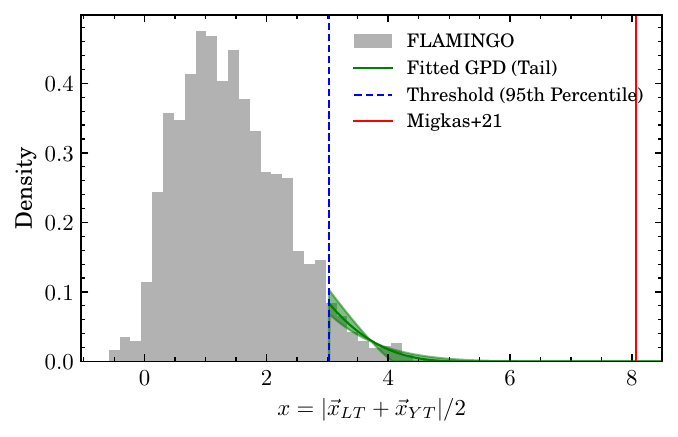}
        \subcaption{No scatter, M21 method}
        \label{fig:evs-joint-m21}
    \end{subfigure}
    \begin{subfigure}[b]{0.48\linewidth}
        \centering
        \includegraphics[width=\linewidth]{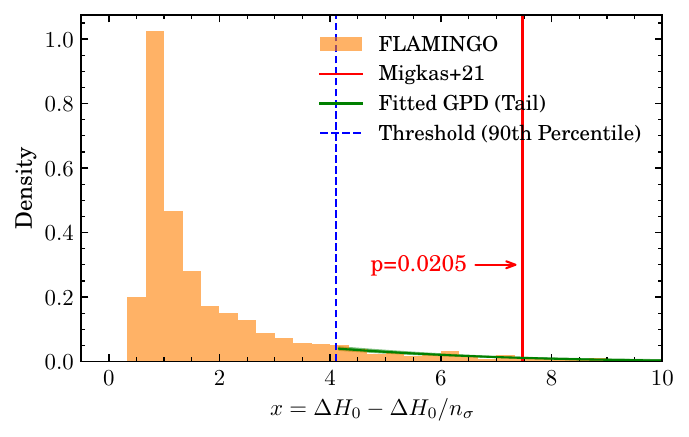}
        \subcaption{Matched scatter, MCMC}
        \label{fig:evs-joint-scatter}
    \end{subfigure}
    \begin{subfigure}[b]{0.48\linewidth}
        \centering
        \includegraphics[width=\linewidth]{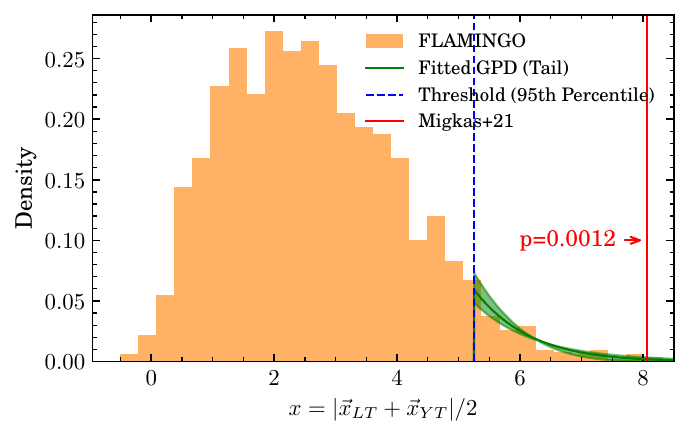}
        \subcaption{Matched scatter, M21 method}
        \label{fig:evs-joint-m21-scatter}
    \end{subfigure}
    \caption{Extreme value statistics analyses of joint $H_0$ anisotropy in the FLAMINGO lightcones using the MCMC (left) and M21 (right) methods without (top) and with (bottom) matched scatter. Grey and orange histograms show the lower-bound projection of all 1728 lightcones without and with scatter, respectively. The red vertical line marks the M21 result under the same projection. The blue dashed line indicates the GPD threshold (90th or 95th percentile), above which the tail is modelled by a GDP, shown in green. {The small shaded green region around the best fit line represents the 16th to 84th percentile region of bootstrapping error of the fit.} Annotated in red is the EVS probability of obtaining the M21 result in FLAMINGO.}
    \label{fig:evs-all-joint}
\end{figure*}

The MCMC joint constraint is shown in Fig.~\ref{fig:h0-result-joint-MCMC}, alongside the M21 result of $\Delta H_0 = 9\%$ at $5.9\sigma$. The EVS analyses are presented in Fig.~\ref{fig:evs-joint} and \ref{fig:evs-joint-scatter}. The M21 point lies well outside the $3\sigma$ region spanned by the lightcones (grey contours). Using projected EVS (Sect.~\ref{sec:method-probability}), we estimate the probability of recovering the M21 result within FLAMINGO lightcones to be $p = 1.5 \times 10^{-4}$, corresponding to a $3.78\sigma$ tension (Fig.~\ref{fig:evs-joint}). When the same test is repeated including injected scatter, the tension reduces to $p = 0.020$ ($2.32\sigma$ with EVS, Fig.~\ref{fig:evs-joint-scatter}). Although the added scatter brings the result visibly closer to M21, the KDE still yields a low $p$-value. As seen in Fig.~\ref{fig:h0-result-joint-MCMC}, high $\Delta H_0$ and $n_\sigma$ are attainable in the MCMC method, but the combination of high $n_\sigma$ and low $\Delta H_0$ remains rare. However, since the M21 MC result was obtained using a slightly different statistical method (isotropic MC samples rather than full-sky MCMC), it is not guaranteed to follow the same 2D distribution. Thus, KDE-based probabilities with MCMC, especially in the tails, should be treated with caution in this comparison.

A joint probability from the M21 method results can also be obtained. Strictly following M21 and creating $10^5$ isotropic samples for each lightcone is computationally unfeasible. Instead, we expand the projection to an average of $\LT$ and $\YT$, which naturally accounts for the potential correlation between the two relations. To incorporate directional alignment, we define a dimension-reduction function $x = |\vec{x}_\mathrm{LT} + \vec{x}_\mathrm{YT}|/2$, where $|\vec{x}_{YX}| = \Delta H_0 - \Delta H_0/n_\sigma$ and the vector direction corresponds to the anisotropy dipole. This reduces the five-dimensional space\footnote{$\Delta H_0$ and $n_\sigma$ for $\LT$ and $\YT$, plus their angular separation.} to a single variable, allowing us to apply the EVS method for probability estimation. The corresponding KDE estimation would require integrating the full five-dimensional space, which is beyond the scope of the method. The two entries in Table~\ref{tab:results-bf} for the Joint M21 analyses using KDE are thus empty.

Without added scatter, the M21 result\footnote{{As a rough estimation, the M21 result is $6.7$ standard deviations away from the median of the lightcones; however, this cannot be taken at face value and translated to a $p$-value because the distribution (Fig.~\ref{fig:evs-joint-m21}) is not Gaussian.}} lies beyond the numerical reach of even the EVS method, as shown in Fig.~\ref{fig:evs-joint-m21}. The corresponding entry is therefore left empty in Table~\ref{tab:results-bf}. When matched scatter is included, the estimated probability becomes $p = 0.0012$ ($3.24\sigma$, EVS, Fig.~\ref{fig:evs-joint-m21-scatter}), indicating a moderate tension with $\Lambda$CDM. This constitutes our primary constraint on the $H_0$ anisotropy scenario: under identical statistical treatment and matched scatter, the M21 anisotropy result combining $\LT$ and $\YT$ differs from the FLAMINGO lightcones at $3.24\sigma$.

In interpreting these estimations, one should note that the joint result in M21 incorporates $110\dash 170$ additional clusters from the ASCA Cluster Catalogue \citep[ACC;][]{horner_x-ray_2001} and an extra scaling relation, $L_\mathrm{BCG}\dash T$, where $L_\mathrm{BCG}$ is the infrared luminosity of the brightest central galaxy per cluster. However, the contribution of $L_\mathrm{BCG}\dash T$ is minor (with individual constraint at $1.9\sigma$ dipole significance and $0.8\sigma$ from mock samples), and the statistical impact of the ACC sample is compensated in our MCMC analyses by using $N = 400$, which better matches the combined data size of 481 for $\LT$ and 373 for $\YT$ in M21. 

In addition, we assess the correlation of the $L_\mathrm{X}$ and $Y_\mathrm{SZ}$ scatter with respect to $T$ in FLAMINGO. We find that the two quantities exhibit a correlation with a Pearson's correlation coefficient $r=0.25$ ($0.31$ with matched scatter; see Appendix~\ref{app:scatter-correlation}). This correlation is weaker than what is found in M21 ($r=0.67$). However, this should not bias our result significantly because in M21 $\LT$ has $\sim 100$ more clusters than $\YT$, meaning $\sim 100$ samples are not common, and therefore the effect of a stronger correlation is mitigated.

\begin{table*}
    \centering
    \caption{Summary of the probability analyses quantifying the tension between the M21 results and the FLAMINGO lightcones using different scaling relations ($\LT$, $\YT$, or their combination), statistical methods (M21 sky scanning or MCMC), and probability estimators (KDE or projected EVS).} 
    \begin{tabular}{ccccc}
    \toprule
       Relation & M21 (KDE) & M21 (projected EVS) & MCMC (KDE) & MCMC (projected EVS) \\
    \midrule  \addlinespace
        $\LT$ & $p=0.021$ & $p=0.025\, (2.25\sigma)$ & $p=0.055$ & $p=0.032\,
        (2.14\sigma)$\\
     \addlinespace
        $\YT$ & $p < 5.8\times 10^{-4}$ & $p=1.8\times10^{-10}\, (6.38\sigma)$ &
        $p < 5.8\times 10^{-4}$ & $p=2.8\times 10^{-4}\, (3.64\sigma)$\\
     \addlinespace
        Joint analysis & -- & -- & $p < 5.8\times 10^{-4}$
        & $p=1.5\times 10^{-4}\,(3.78\sigma)$\\
    \midrule  \addlinespace
        $\LT$ (scatter) & $p=0.33$ & $p=0.23\,(1.20\sigma)$ & $p < 5.8\times
        10^{-4}$
        & $p=0.085\,(1.72\sigma)$\\
     \addlinespace
        $\YT$ (scatter) & $p < 5.8\times 10^{-4}$ 
        & $p=0.0013\,(3.22\sigma)$ & $0.0014$ & $p=0.0049\,(2.82\sigma)$\\
     \addlinespace
        Joint analysis (scatter) & $\dash$ &
        $p=0.0012\,(3.24\sigma)$ & $p<5.8\times 10^{-4}$ &
        $p=0.020\,(2.32\sigma)$\\
    \bottomrule
    \end{tabular}
    \tablefoot{The $p$-values denote the probability that a randomly placed observer in a $\Lambda$CDM universe would measure an anisotropy equal to or more extreme than that seen in M21, with equivalent (1D) Gaussian significance given in parentheses where applicable. `Scatter' refers to the mock samples with matched M21-level instrumental uncertainties (see Sect.~\ref{sec:matched-scatter}). Dashes indicate configurations beyond the scope of the probability estimators.}
    \label{tab:h0-$p$-value-summary}
\end{table*}

\subsection{Bulk flow constraints} \label{sec:results-bf}

\begin{figure*}
    \centering
    \begin{subfigure}[b]{0.48\linewidth}
        \centering
        \includegraphics[width=\linewidth]{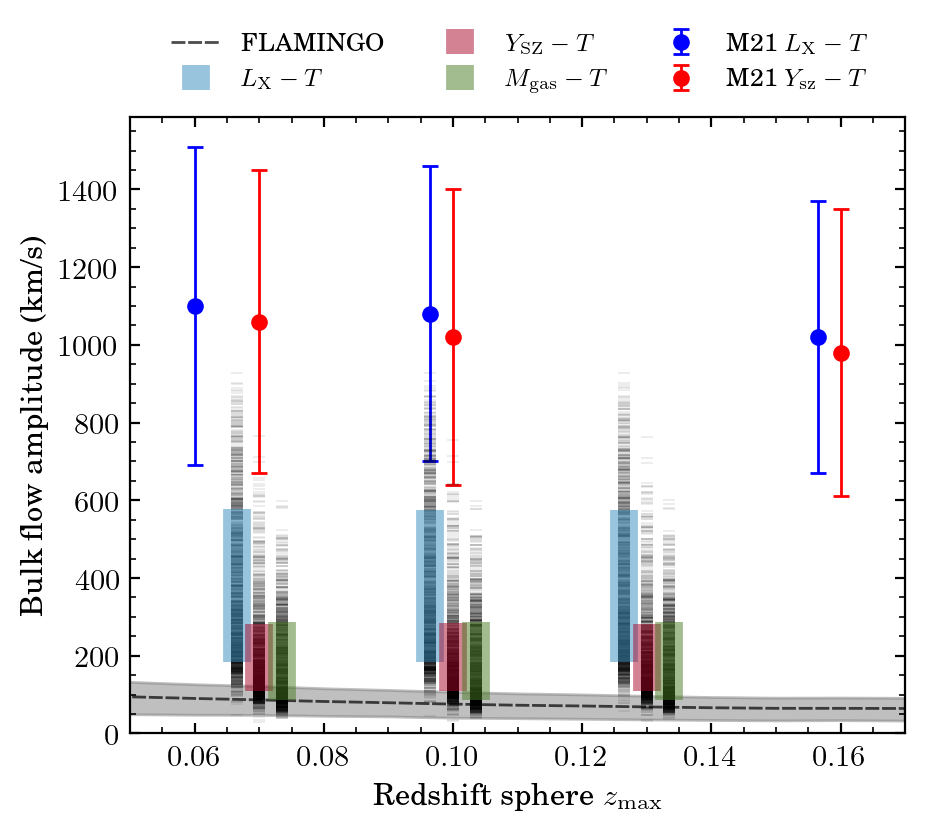}
        \subcaption{No scatter}
    \end{subfigure}
    \begin{subfigure}[b]{0.48\linewidth}
        \centering
        \includegraphics[width=\linewidth]{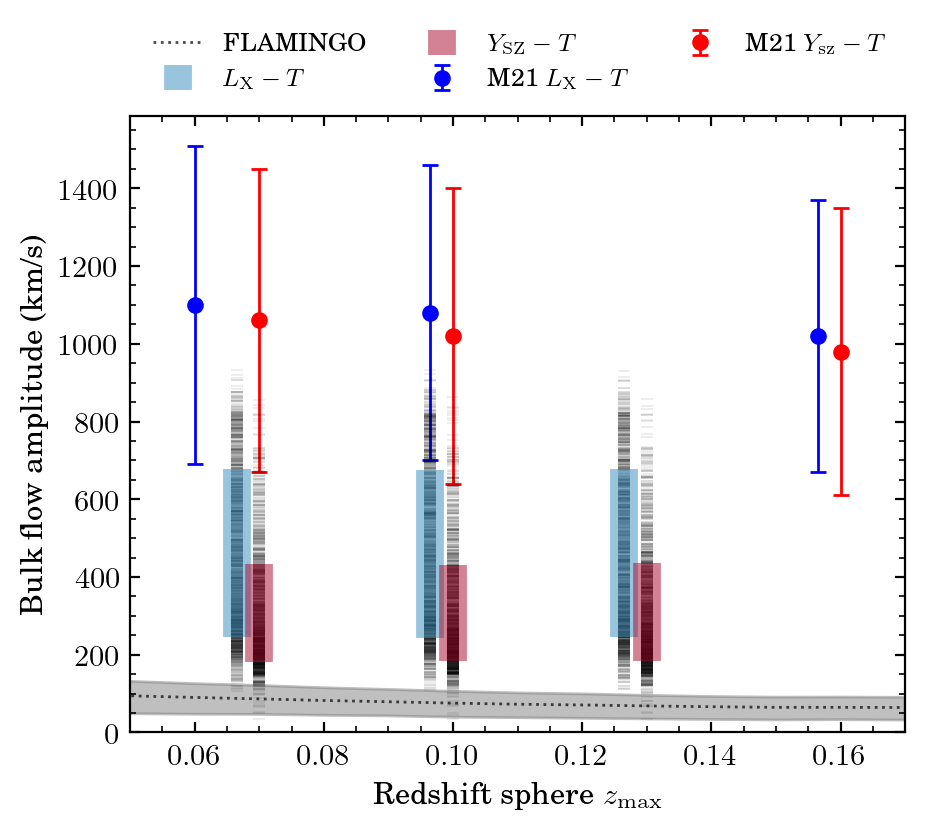}
        \subcaption{Matched scatter}
    \end{subfigure}
    \caption{The MCMC bulk flow amplitude constraints for $z < 0.07$, $z < 0.10$, and $z < 0.13$ shown for the $\LT$, $\YT$, and $\MT$ relations. The right panel includes injected scatter and the left panel does not. Black dashes represent the results from individual lightcones, while the thick coloured bars indicate the 16th to 84th percentile range across the lightcones. For comparison, the M21 results for $\LT$ and $\YT$ are shown with their original error bars, corresponding to redshift spheres of $z < 0.06$, $z < 0.07$, $z < 0.10$, and $z < 0.16$. To avoid visual overlap, the points for $\LT$ and $\MT$ are slightly offset along the $x$-axis. The black dashed line represents the median true average bulk motion of the clusters, with the grey shaded region indicating the 16th to 84th percentile range of the lightcones.}
    \label{fig:results-bf}
\end{figure*}

\begin{figure}
    \centering
    \includegraphics[width=\linewidth]{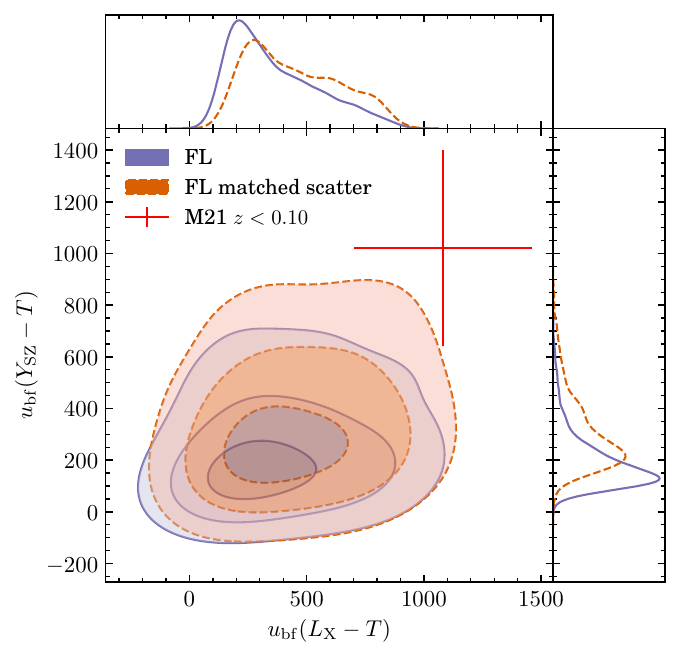}
    \caption{Distribution of the $\LT$ and $\YT$ bulk flow constraints at $z<0.10$ along with the M21 result at the same redshift with its error bars (red). The orange contour includes injected scatter, while the purple does not. Contours of different shades give the 39\%, 86\%, and 98.9\% regions.}
    \label{fig:result-bfmc_hist_joint_LTYT}
\end{figure}

\begin{table*}
\centering
\caption{Summary of bulk flow results from the FLAMINGO lightcones for different scaling relations and redshift ranges in direct comparison with the observed values from M21.}
\label{tab:results-bf}
\footnotesize
\begin{tabular}{cccccccc}
\toprule
Relation & \makecell[c]{Redshift\\Range} & \makecell[c]{No scatter\\ $u_\mathrm{bf}$
($\SI{}{km.s^{-1}}$)} & \makecell[c]{Matched scatter\\$u_\mathrm{bf}$
($\SI{}{km.s^{-1}}$)} & \makecell[c]{FLAMINGO\\ $|\langle \vb{v} \rangle|$
($\SI{}{km.s^{-1}}$)} & \makecell[c]{M21 observed\\ $u_\mathrm{bf}$
($\SI{}{km.s^{-1}}$)} & $p$-value of M21 $(\sigma)$ & \makecell[c]{$p$-value of
M21 $(\sigma)$\\matched scatter}\\
\midrule
\multirow{3}{*}{$\LT$} & $z < 0.07$ & $315^{+262}_{-131}$ & $414^{+263}_{-169}$
& $86^{+40}_{-35}$ & $1100_{-410}^{+410}$ & \tablefootmark{a}$0.0067\,(2.71\sigma)$ &
\tablefootmark{a}$0.0081\,(2.65\sigma)$\\
    \addlinespace
     & $z < 0.10$ & $315^{+260}_{-131}$ & $415^{+260}_{-172}$ & $75^{+36}_{-30}$ &
     $1080^{+380}_{-380}$ & $0.0082\,(2.64\sigma)$ & $0.0067\,(2.72\sigma)$\\
    \addlinespace
     & $z < 0.13$ & $313^{+263}_{-130}$ & $417^{+261}_{-172}$ & $68^{+34}_{-27}$
     & -- & -- & --\\
    \addlinespace
\midrule
    \addlinespace
    \multirow{3}{*}{$\YT$} & $z < 0.07$ & $164^{+119}_{-56}$ &
    $264^{+170}_{-83}$ & $86^{+40}_{-35}$ & $1060^{+390}_{-390}$ & $3.2\times
    10^{-6}\,(4.66\sigma)$ & $5.8\times 10^{-4}\,(3.44\sigma)$\\
    \addlinespace
      & $z < 0.10$ & $164^{+119}_{-56}$ & $264^{+166}_{-81}$ & $75^{+36}_{-30}$
      & $1020^{+380}_{-380}$ & $3.5\times 10^{-6}\,(4.64\sigma)$ &
      $0.0011\,(3.26\sigma)$\\
    \addlinespace
      & $z < 0.13$ & $164^{+117}_{-56}$ & $264^{+172}_{-82}$ & $68^{+34}_{-27}$
      & -- & -- & --\\
    \addlinespace
\midrule
    \addlinespace
    \multirow{3}{*}{$\MT$} & $z < 0.07$ & $158^{+129}_{-73}$ & -- & $86^{+40}_{-35}$
    & -- & -- & --\\
    \addlinespace
    & $z < 0.10$ & $158^{+130}_{-73}$ & -- & $75^{+36}_{-30}$ & -- & -- & --\\
    \addlinespace
    & $z < 0.13$ & $158^{+130}_{-72}$ & -- & $68^{+34}_{-27}$ & -- & -- & --\\
    \addlinespace
\midrule
    \addlinespace
    \multirow{2}{*}{\makecell{$\LT+\YT$\\Combined}} & $z < 0.07$ & -- & -- & -- &
    -- & \tablefootmark{a}$5.0\times10^{-5}\,(4.05\sigma)$ & \tablefootmark{a}$3.6\times10^{-4}\,(3.57\sigma)$\\
    \addlinespace
    & $z < 0.10$ & -- & -- & -- & -- & $4.5\times10^{-5}\,(4.08\sigma)$ &
    $3.8\times10^{-4}\,(3.56\sigma)$\\
    \addlinespace
\bottomrule
\end{tabular}
\tablefoot{The columns show the fitted bulk flow amplitudes, $u_\mathrm{bf}$ (with and without matched scatter), the actual bulk peculiar velocity in FLAMINGO ($|\langle \vb{v} \rangle|$), the observed M21 values, and the resulting $p$-values (with Gaussian-equivalent significance in parentheses). Results are reported for three redshift shells and for the combined $\LT+\YT$ analysis. Dashes indicate configurations not evaluated.}
\raggedright 
\tablefoottext{a}{Precisely, the $\LT$ bulk flow in M21 is
measured in $z<0.06$; however, this difference is negligible.}
\end{table*}

Assuming isotropic Hubble expansion, we estimate the bulk flow using the MCMC implementation described in Sect.~\ref{sec:method-bulk-flow}. The bulk flow is constrained within three $z$ spheres: $z<0.07$, $z<0.10$, and $z<0.13$. These are chosen to balance the need for a sufficient number of clusters with maintaining a relatively uniform redshift distribution within each sphere. Additionally, they allow for direct comparisons with M21.

The bulk flow results are presented in Fig.~\ref{fig:results-bf} and summarised in Table~\ref{tab:results-bf}, alongside the M21 measurements and the true FLAMINGO bulk flow. The FLAMINGO bulk flow is estimated by the average of cluster velocities of our sample, $|\langle\vb{v}\rangle| = |\sum_i \vb{v}_i| / N$. As expected, we find that the true bulk flows are $\lesssim \SI{100}{km.s^{-1}}$ across our redshift range, consistent with $\Lambda$CDM predictions \citep{qin_cosmic_2021, watkins_analysing_2023}. Using the $\LT$, $\YT$, and $\MT$ relations, one finds a $\approx 315_{-131}^{+262}\,\SI{}{km.s^{-1}}$, $164_{-56}^{+119}\,\SI{}{km.s^{-1}}$, and $158_{-73}^{+130}\,\SI{}{km.s^{-1}}$ bulk flow respectively, in all three redshift spheres. It is evident that $\LT$ returns the most overestimated bulk flow compared to the true FLAMINGO value (black dashed line in Fig.~\ref{fig:results-bf}) due to its larger scatter. On the other hand, the low-scatter $\YT$ and $\MT$ relations return values much closer to the true underlying bulk flow ($<1.5\sigma$ difference). When scatter is matched to M21, the recovered bulk flow increases to about $415^{+260}_{-172}\,\SI{}{km.s^{-1}}$ for $\LT$ and $264^{+166}_{-81}\,\SI{}{km.s^{-1}}$ for $\YT$ (at $z<0.10$, similar at other redshift shells), amplifying the overestimation compared to the true value. The systematic overestimation of the bulk flow by all three relations and the its increase with mock scatter indicate that the bulk flow constraints also depend on the scatter in the cluster scaling relations. 

Compared to M21, we find no prominent bulk flow exceeding the level of M21 ($\sim \SI{900}{km.s^{-1}}$). However, the dispersion of our bulk flow constraints and the observational uncertainties of M21 are both large. We estimate the statistical significance of this discrepancy while accounting for the large uncertainties by adopting the same EVS-based projection method used in the $H_0$ analysis. We define the effective 1D variable $x = u_\mathrm{bf} - \sigma_\mathrm{low}$, where $\sigma_\mathrm{low}$ is the lower $1\sigma$ uncertainty of each lightcone fit, and assess the probability of recovering the M21 value under this projection using the EVS method (Sect.~\ref{sec:method-probability}).

The resulting $p$-values at $z<0.07$ and $z<0.10$ are given in Table~\ref{tab:results-bf} (corresponding EVS plots are included in Appendix~\ref{app:about-evs}). At $z<0.07$, using $\LT$, the probability of obtaining the M21-level bulk flow from an isotropic lightcone is $p = 0.0067\,(2.71\sigma)$ without scatter and $p = 0.0081\,(2.65\sigma)$ with matched scatter. For $\YT$, the tension is stronger: $p = 3.2 \times 10^{-6}\,(4.66\sigma)$ (no scatter) and $p = 5.8 \times 10^{-4}\,(3.44\sigma)$ (with scatter). The $z<0.10$ results are similar, yielding $2.72\sigma$ ($\LT$) and $3.26\sigma$ ($\YT$) with scatter.

A joint $\LT+\YT$ bulk flow distribution is shown in Fig.~\ref{fig:result-bfmc_hist_joint_LTYT}. Compared to the $H_0$ case, increasing the scatter has a smaller effect. This implies that bulk flow amplitudes are less sensitive to scaling relation scatter in our implementation. We estimate the joint significance by taking the average bulk flow vector amplitude as $x = |\vec{x}_\mathrm{LT} + \vec{x}_\mathrm{YT}|/2$, following the same prescription as in the $H_0$ analysis (Sect.~\ref{sec:results-h0-joint-analysis}). This results in a final tension of $p = 3.6 \times 10^{-4}\,(3.57\sigma)$ at $z<0.07$ and $p = 3.8 \times 10^{-4}\,(3.56\sigma)$ at $z<0.10$, under matched scatter. Without scatter injection, the tension is only slightly higher at $4.05\sigma$ ($z<0.07$) and $4.08\sigma$ ($z<0.10$). While no single lightcone reaches $u_\mathrm{bf} \sim \SI{900}{km.s^{-1}}$, such values remain rare even after accounting for scatter. The resulting tension with $\Lambda$CDM, at the $\sim 3.6\sigma$ level, exceeds that of the $H_0$ anisotropy ($2.3$--$3.2\sigma$). This may point towards a more localised, low-redshift departure from $\Lambda$CDM predictions.

In addition, we notice that the bulk flow tension seems to be less affected by scatter compared to the $H_0$ anisotropy. Here the tension only drops by $0.4\sigma$ with matched scatter, whereas the joint $H_0$ tension dropped $>1.4\sigma$ with MCMC and more with the M21 method. 

We note that for the bulk flow, the M21 observational analysis employed a $\chi^2$ minimisation over the full sky, while we adopt a full-likelihood  maximisation via MCMC. In addition, M21 estimated uncertainties using bootstrap resampling, whereas we derive them from the posterior distributions of the MCMC samples. These methodological differences may introduce variations in the estimated uncertainties. However, as $\chi^2$ minimisation is mathematically equivalent to maximising the likelihood under Gaussian assumptions, the impact on the central values is expected to be minimal. If anything, the use of MCMC could lead to slightly narrower uncertainty estimates. Since we evaluate the statistical tension using the lower $1\sigma$ bound of each lightcone's  distribution, any underestimation of uncertainties would in fact strengthen the inferred tension with $\Lambda$CDM.

\section{Discussion}\label{sec:discussion}

\subsection{Cosmological implications}\label{sec:discussion-cosmological-implication}

By generating 1728 lightcones in the FLAMINGO $(\SI{2.8}{Gpc})^3$ box, we find that the M21 results are unlikely to occur within a $\Lambda$CDM universe. In joint analyses of the $\LT$ and $\YT$ relations, the amplitude of $H_0$ variation and the significance reported by M21 lie beyond the range of simulated lightcones. We estimate the overall tension between the M21 findings and $\Lambda$CDM to be $\sim2.3$--$3.2\sigma$, depending on the exact statistical method and probability estimator. The alternative explanation based on a coherent bulk flow also shows significant tension: at $z<0.1$, the M21-level bulk flow constitutes a $3.6\sigma$ deviation. Notably, none of our 1728 lightcones produce such a large bulk flow, even when observational scatter is included. The fact that the statistical tension increases for bulk flow model $z$ compared to $H_0$ suggests that the observed anisotropy is at a higher discrepancy locally at $z<0.1$.

Overall, these results suggest that both scenarios proposed in M21 to explain their findings ($H_0$ anisotropy or a significant bulk flow) are in tension with $\Lambda$CDM and, although substantially reduced, cannot be fully attributed to statistical noise or cosmic variance.Assuming no major undiscovered systematic biases as yet in the M21 results or the FLAMINGO simulations, this work implies that either the $\Lambda$CDM model needs modifications at $z < 0.2$ or that we inhabit a statistically rare region of the Universe. Any of the two scenarios can potentially have significant implications for other low$-z$ cosmological tensions, such as the Hubble tension. Local $H_0$ measurements typically assume isotropy in the expansion rate and $\Lambda$CDM-predicted bulk flows. If either assumption fails, as our findings suggest may be the case, it could systematically bias the inferred $H_0$ values \citep{migkas_galaxy_2025}. This opens the possibility that part of the Hubble tension arises from local anisotropies rather than global physics.

Another key finding with direct cosmological relevance is the role of scatter in cluster scaling relations. We show that the observed anisotropy amplitude is boosted by scatter in $\LT$ and $\YT$, even when the input cosmology is isotropic. Tighter relations, such as $\MT$, exhibit lower bias and are therefore better suited for future anisotropy studies. For example, with matched scatter, we find that FLAMINGO lightcones overestimate $\Delta H_0$ by $\sim(7$--$8)\%$ when $\LT$ or $\YT$ are used alone (see Fig.~\ref{fig:h0-result}). This implies that the $\Delta H_0 = 9\%$ reported in M21 may correspond to a true underlying anisotropy of only $\sim 3\%$\footnote{The M21 detection was $\Delta H_0 = 14\%$ at $4.1\sigma$ using $\YT$ and $8.7\%$ at $2.8\sigma$ using $\LT$. The $\YT$ detection likely retains an amplitude above $1\%$.} or so. Properly addressing this bias would require injecting an artificial anisotropy signal into simulations and comparing the measured with the input values (see Sect.~\ref{sec:injecting-signal}). If a strong signal exists, the bias from noise may be smaller. Future work will provide a more conclusive assessment.

{Finally, to aid the interpretation of our results, we summarise the important assumptions that we adopt. First and most importantly, we assume that the shape of the anisotropies is dipole-like and study only dipole anisotropies, either dipole $H_0$ variations or a single bulk flow component. We do this because we focus on understanding the results of M21, which found a dipole-shaped anisotropy. Higher order anisotropies are beyond the scope of this work. Second, our interpretation that translates the probability of an anisotropy in the simulations to the probability of that in a $\Lambda$CDM universe assumes that the FLAMINGO simulation (with its physical modelling and numerical implementations) is an accurate realisation of the $\Lambda$CDM cosmology. And third, the scatter injection method as described in Sect.~\ref{sec:matched-scatter} assumes that the scatter is mass independent and with a log-normal distribution.}

\subsection{Origin of apparent anisotropies: Statistical noise or peculiar velocities}

We now discuss the possible origins of the apparent anisotropies observed in the FLAMINGO lightcones. One key contributor is bulk flow. Coherent peculiar motions can introduce real anisotropy signals that manifest consistently across different scaling relations. As shown in Fig.~\ref{fig:results-bf}, FLAMINGO reproduces the expected low-amplitude large-scale bulk flows of $\Lambda$CDM. Such motions are predicted to induce apparent $H_0$ anisotropies at the $\sim 3\%$ level \citep{migkas_galaxy_2025}, which is close to the amplitudes found in our simulated lightcones.

A second major source is statistical noise, particularly random upscattering of $\Delta H_0$ in individual lightcones. In this scenario, different scaling relations would yield weakly correlated anisotropies in the same lightcone. Especially ones that are less correlated in scatter (Appendix~\ref{app:scatter-correlation}). The two sources can be disentangled by examining the degree of alignment between anisotropies inferred from different relations. Distinguishing between bulk flow and statistical noise is crucial for guiding future studies of cosmic anisotropy and for interpreting the significance of observed signals.

\subsubsection{Directional correlation between different
relations}\label{subsubsec:directional-correlation-between-different-relations}

To investigate whether the apparent anisotropies arise from a genuine bulk flow, we examine whether different scaling relations exhibit their maximum dipole in similar directions. We calculate the angular separation between the directions of the strongest dipole detected in each relation. Our results indicate that most relation pairs do not show significant directional alignment. The only pair with a mild angular correlation is $\LT$ and $\MT$, whose dipoles peak around a separation angle of $\sim 20^\degree$. By contrast, $\YT$ and $\MT$ show only weak alignment, and $\LT$ and $\YT$ exhibit no consistent directional correlation. We also test whether a larger $\Delta H_0$ is associated with better directional agreement, i.e. smaller angular separation, but find no such correlation.

These findings suggest that the apparent anisotropies in the simulated lightcones are most likely predominantly driven by statistical noise rather than by peculiar velocity fields consistent with $\Lambda$CDM. This interpretation is further supported by the scatter correlation patterns (Appendix~\ref{app:scatter-correlation}), where relations that align in scatter also tend to align in dipole direction. (For a visualisation of the angular separation distributions and their cross-correlations, see Appendix~\ref{app:angular-separation}.)

\subsubsection{Effect of peculiar velocities in apparent anisotropies}

Peculiar velocities contribute to anisotropy signals in two ways: random motions increase the scatter in scaling relations, while coherent bulk flows can mimic an $H_0$ dipole. To assess their impact, we repeat the anisotropy analyses using cosmological redshifts $z_\mathrm{cos}$ instead of the observed redshifts $z_\mathrm{obs}$ (see Sect.~\ref{sec:lightcone-construction}). In this setup, clusters are effectively placed at their Hubble-flow distances, completely removing the influence of peculiar velocities.

We find that this replacement leads to a mild reduction in anisotropy amplitudes. For $\LT$, the median $\Delta H_0$ decreases from $3.94\%$ to $3.35\%$; for $\YT$, from $2.71\%$ to $2.24\%$; and for $\MT$, from $1.50\%$ to $1.26\%$. Hence, peculiar velocities—bulk flows included—account for only $\sim15$--$17\%$ of the apparent anisotropy, with the remaining contribution attributed to statistical noise.

A similar trend is seen in the bulk flow fits. For $\LT$, the amplitude decreases slightly from $\SI{313}{km.s^{-1}}$ to $\SI{303}{km.s^{-1}}$ across all redshift bins. The $\MT$ relation shows a similar drop from $\SI{161}{km.s^{-1}}$ to $\SI{151}{km.s^{-1}}$, and $\YT$ exhibits a stronger decline from $\SI{165}{km.s^{-1}}$ to $\SI{135}{km.s^{-1}}$. Even when peculiar velocities are fully removed, the reduction in signal remains below $\sim18\%$, reinforcing the interpretation that statistical noise dominates the recovered anisotropies.  This suggests that our bulk flow analysis is primarily dominated by noise when the underlying bulk flow amplitude is $u_\mathrm{bf} \lesssim \SI{100}{km.s^{-1}}$ at $z \sim 0.10$ (Table~\ref{tab:results-bf}).

\subsection{Effects of scaling relation scatter in apparent anisotropies}\label{sec:effect-of-scatter}

We find that the scatter in scaling relations directly affects the apparent anisotropies. Fig.~\ref{fig:h0-result-joint-MCMC} shows that injecting scatter boosts $\Delta H_0$ while leaving $n_\sigma$ nearly unchanged. This behaviour holds across individual relations and with the M21 method (Fig.~\ref{fig:h0-result}). The reason is that scatter amplifies both the anisotropy amplitude and its uncertainty, keeping the statistical significance approximately stable. As a result, $\Delta H_0$ is more sensitive to observational systematics than $n_\sigma$, and our projection-based methods may not fully capture this effect, potentially underestimating the true tension.

Importantly, it is not the total scatter of the relation itself but the scatter in $\Delta H_0$ that matters. For instance, with injected scatter, the $\YT$ relation yields a comparable $\Delta H_0$ to $\LT$ under the M21 method (Fig.~\ref{fig:h0-result-chi2-scatter}), despite having $\sim 25\%$ lower $\sigma_\mathrm{tot}$ (Table~\ref{tab:scaling-relation-behaviours}). In this sense, the MCMC method is more robust to noise, as its contour shapes remain less disturbed and low-scatter relations give lower $\Delta H_0$ consistently (Fig.~\ref{fig:h0-result-MCMC-scatter} and \ref{fig:h0-result-joint-MCMC}). The $\MT$ relation, which shows the lowest $H_0$ scatter, yields the most stable results. Future studies should prioritise low-scatter relations to reduce bias, and apply caution when interpreting anisotropy amplitudes from high-scatter relations. 

Finally, we note that although the simulated scatter is generally matched to the observational values, there remains uncertainty in the scatter, the scatter may still be slightly underestimated if the uncertainties are not well matched. For the $\LT$ ($\YT$) relation, the typical fitting uncertainty in the intrinsic scatter is $\sim0.015$ ($0.012$) dex\footnote{The values differ from Table~\ref{tab:scaling-relation-behaviours} because here we consider the uncertainty in each individual lightcone, rather than the distribution across all lightcones, which is what is reported in Table~\ref{tab:scaling-relation-behaviours}.} in our lightcones, which closely matches the M21 uncertainties of $0.016$ and $0.013$ dex, respectively. However, we find a slight underestimation in the uncertainty of the total scatter, $0.015$ dex for $\LT$ and $0.010$ dex for $\YT$, compared to $0.018$ and $0.015$ dex in M21. This could result in a marginal underestimation of the scatter, even after our scatter-matching. Nonetheless, the effect should be small. The bootstrapped scatter distributions in both M21 and FLAMINGO are approximately Gaussian and do not exhibit strong tails. Anomalies caused by a few strongly scattered clusters would therefore be reproduced in the lightcones. Moreover, it was carefully examined in both M20 and M21 that there is no strong outlier causing the observed anisotropy, further reducing this effect.

\subsection{{Effects of the highest X-ray concentration selection}}\label{sec:effect-of-xray-concentration}

As we noted in Sect.~\ref{sec:sample-selection}, for the same selection criteria, FLAMINGO yields more clusters than the observed M21, but with a lower $c_\text{X}$ distribution. The origin of this discrepancy requires further investigation as we discuss in Appendix.~\ref{app:xray-concentration}. For our analysis, we adopted a temporary solution by selecting the highest $c_\text{X}$ clusters from the simulated samples. Since the selection is performed independently of the sky position of clusters and should have no direct impact on our results. Nevertheless, there might be secondary systematic effects that can potentially bias our results (e.g. via affecting the intrinsic scatter of scaling relation).

To check whether our final results are robust against the applied $c_\text{X}$ selection, we repeat the MCMC tests with random cluster selection independently of the $c_\text{X}$ values. Specifically, after the flux and zone-of-avoidance selections (described in Sect.~\ref{sec:sample-selection}), for each lightcone, we randomly drew 313 (400) clusters for the M21 (MCMC) statistical approach, out of the $700 \dash 800$ remaining clusters. Using the random cluster selection in each lightcone, we quantify the tension between the M21 findings and $\Lambda$CDM to be $2.27\sigma$ for $\LT$, $3.38\sigma$ for $\YT$, and $3.42\sigma$ when the two relations combined. The $\LT$ tension increased by $0.1\sigma$ while the $\YT$ tension decreased by $0.2\sigma$, and the combined tension slightly reduces by $\sim 0.4\sigma$. With matched scatter, the $\LT$ tension is approximately unchanged at $1.74\sigma$, while the $\YT$ tension is at $2.60\sigma$, slightly ($0.2\sigma$) lower than when the $c_\mathrm{X}$-bases selection was used. When combined in the joint probability, the overall tension with matched scatter is $2.32\sigma$, which is essentially the same with the default analysis ($2.31\sigma$, Table~\ref{tab:h0-$p$-value-summary}). When modelled as a bulk flow and including the injected scatter, we find a joint (combining $\LT$ and $\YT$, as described in Sect.~\ref{sec:results-bf}) $3.63\sigma$ ($3.56\sigma$) for $z<0.07$ ($z<0.10$), similar to the default results shown in Table~\ref{tab:results-bf}. Without matched scatter, the joint tension exceeds $4\sigma$, and the corresponding $p$-value is so low that even the EVS method cannot reliably estimate it.

{Based on the above tests, we conclude that although the $c_\mathrm{X}$ discrepancy between observational data and the FLAMINGO simulation warrants further investigation, it does not noticeably impact the main conclusions of this work. The $\gtrsim 3\sigma$ tension between our simulation lightcones and M21 is neither caused by nor sensitive to our top $c_\mathrm{X}$ selection during sample selection (Sect.~\ref{sec:sample-selection}).}

\subsection{Comparing the M21 method to MCMC}\label{subsec:method-compare}

Constraints on cosmic anisotropy from galaxy clusters depend sensitively on the statistical method used, underscoring the importance of robust approaches with minimal noise. We have presented results from both the M21 and MCMC methods separately, and now compare them directly by correlating outcomes for the same lightcones. In the absence of scatter, $\Delta H_0$ values from M21 and MCMC show only weak correlation when using the same scaling relation, with Pearson coefficients of $\sim 0.5$ for $\LT$, $\YT$, and $\MT$. A similarly weak correlation is found in their respective $n_\sigma$ values ($r \lesssim 0.5$; see Appendix~\ref{app:m21-vs-mcmc}), consistent with the interpretation that apparent anisotropies in the isotropic mock lightcones are dominated by statistical noise rather than bulk flows.

Without a true cosmic anisotropy or a bulk flow mimicking one, the MCMC method yields less noisy constraints, with narrower distributions in both $\Delta H_0$ and $n_\sigma$ (Fig.~\ref{fig:h0-result}) across all three scaling relations and more stable to injected noise (Sect.~\ref{sec:effect-of-scatter}). It shows better angular consistency between dipoles inferred from different relations (see Sect.~\ref{subsubsec:directional-correlation-between-different-relations}). Additionally, the MCMC method is computationally more efficient, more versatile, and can be extended to non-dipole anisotropies with relative ease.

These advantages, however, are not sufficient to conclude that MCMC outperforms the M21 method in all contexts. First, the MCMC method produces significantly more outliers, which strongly boosts any probability estimation (e.g. EVS) based on the M21 distribution. In future observations, this may increase the risk of reporting statistically significant anisotropies from data that are otherwise consistent with isotropy. While comparison to simulations or other mock samples can mitigate this effect, the heavy-tailed nature of the MCMC result remains challenging to model robustly. Second, our implementation of the MCMC method corresponds to applying the M21 approach with a $90\degree$ cone. Using larger cones tends to smooth out both real signals and statistical fluctuations, lowering both $\Delta H_0$ and $n_\sigma$ (see M20, their Table~2). Third, part of the reduced noise in MCMC may be attributed to the larger cluster sample size: 400 clusters for MCMC versus 313 for M21. Nonetheless, the MCMC-based method offers a promising avenue for future observational applications.

It is worth emphasising that the two methods differ in more than just sampling techniques. The MCMC method uses the full likelihood rather than $\chi^2$, MCMC sampling instead of bootstrap resampling, full-sky fitting rather than cone scanning, and forward modelling rather than a direct $A$ to $H_0$ translation. Any difference in performance must therefore be understood in the context of these broader methodological distinctions.

\subsection{Future prospects}

\subsubsection{Other cluster catalogues}

While the statistical significance of the cosmic dipole reported in M21 is lowered after improved treatment of cosmic variance, a notable tension with $\Lambda$CDM remains at $z\lesssim 0.1$ ($3.6\sigma$). A genuine cosmological signal should appear consistently across datasets, independent of scaling relation, selection function, or instrument. To mitigate potential systematics and selection biases, it is essential to repeat such analyses on independent cluster samples with alternative selection methods (e.g. SZ selection), aided by dedicated mock catalogue studies. We plan to apply our approach to the eROSITA first All-Sky Survey (eRASS1), which comprises over 12000 clusters across $z = 0.01$--$1.2$ \citep{bulbul_srgerosita_2024}. Other catalogues, including X-CLASS \citep{koulouridis_x-class_2021} and XCS \citep{mehrtens_xmm_2012, giles_xmm_2022}, also provide promising avenues for future investigation. A consistent detection of a dipole across such independent datasets would offer a robust test of any large-scale anisotropy of cosmological origin.

\subsubsection{Improvement on lightcone independence}

Our 1728 lightcones do not represent entirely 1728 realisations of the $\Lambda$CDM cosmology. While each lightcone places the observer at a different location and the FLAMINGO simulation spans an unprecedented scale, the overlap between lightcones is unavoidable. Neighbouring observers are $\sim \SI{230}{cMpc}$ apart. As a result, similar structures will appear across neighbouring lightcones at $z \gtrsim 0.05$. The effect on our analysis is kept to a minimum by random rotations and zone-of-avoidance masking (Sect.~\ref{sec:sample-selection}). For each lightcone, a different random galactic plane is masked to reduce repetitive structures. Future analyses can look to address this limitation by developing better techniques for building the samples. For example, using the dense sphere packing technique instead of a cubic grid that we use, the number of independent lightcones can increase by 40\% under the same volume. To further increase the statistical power, larger simulation volumes are needed.

\subsubsection{Response to injected anisotropy}\label{sec:injecting-signal}

All the tests in this work were performed on datasets with no or small (e.g. from bulk flows) signals. We have concluded that the M21 findings are rare if only noise is present. It is then important to study, if exists, whether the amplitude of such an anisotropy is overestimated in M21. In this work, we showed how statistical noise can cause a spurious anisotropy signal. It is likely the same mechanism could boost up an existing signal. Future work should include controlled signal injection to assess the ability of this methodology to recover known input anisotropies. Such tests would also allow for direct comparison with M21 and newer observational results and provide a more systematic validation of the use of cluster scaling relations in probing cosmic anisotropy.

\section{Conclusion}\label{sec:conclusion}

In this work, we have quantified the tension between previously observed cosmic anisotropies using galaxy clusters and the $\Lambda$CDM model using the fully hydrodynamic cosmological FLAMINGO simulations. We created 1728 isotropic lightcones and analysed cluster scaling relations $\LT$, $\YT$, and $\MT$ using two independent statistical approaches: one following M21 and another using MCMC inference with full likelihood. We also increased the intrinsic scatter of the cluster scaling relations and assigned instrumental uncertainties that match M21 data. We have provided a robust assessment of cosmic variance, statistical noise, and systematic biases inherent in the observational findings.

Our analysis reveals that the observed 9\% $H_0$ dipole detected by M21 is rare within a $\Lambda$CDM framework. {For the first test using the original FLAMINGO data (without matched scatter), M21 is at a strong tension with the $\Lambda$CDM simulation. Using the same methodology with $\LT$ and $\YT$, the results differ by 6.7$\sigma$ between M21 and the simulation lightcones. Using the MCMC method yields $3.8\sigma$ ($4.0\sigma$) when the anisotropy is modelled as an $H_0$ variation (bulk flow). However, scatter in the scaling relations plays a significant role in the results, which the simulations underestimate by $\sim 25\%$. Therefore, we also performed tests with scatter injected.} 

{When the intrinsic scatter is matched to observations and mock instrumental uncertainties are assigned,} the joint analysis of the $\LT$ and $\YT$ scaling relations shows tension between data and the standard cosmological model at $2.3 \dash 3.6\sigma$. Repeating the exact M21 method, the M21 finding is $3.2\sigma$ more anisotropic than the FLAMINGO lightcones. Additionally, we examined the bulk flow hypothesis as an alternative explanation for the observed anisotropies. While the simulated lightcones exhibit $\Lambda$CDM-like bulk flows, no bulk flow motion comparable to the $\SI{900}{km.s^{-1}}$ flow observed by M21 was detected in our 1728 lightcones. A $3.6\sigma$ tension between the M21 bulk flow and FLAMINGO lightcones was found at $z<0.07$ and $z<0.10$. This indicates that the $\Lambda$CDM predictions are less compatible with the M21 cluster data at low redshift. Although this finding does not alleviate the dipole found in M21, the statistical tension with $\Lambda$CDM is significantly reduced from $5.4\sigma$ {(M21)} to $3.2\sigma$ {(this work)}.

We find that the amplitude of the detected anisotropies in FLAMINGO is dominated by statistical noise in the cluster scaling relations at the $>80\%$ level, while the remaining $<20\%$ is attributed to peculiar velocities (bulk flows included). We have shown that in the case of dominant statistical noise, different scaling relations yield anisotropy in different directions with no strong correlation between them. This has important implications for future studies, as by utilising independent relations and cluster samples, one can distinguish statistical noise from true cosmological effects by looking at the correlation of the detected dipole directions. The fact that tight direction correlation is very rare also stresses that the M21 finding with three relations and two independent catalogues all pointing to a $< 10\degree$ cone is unlikely to be a coincidence. Last but not least, we find that the inferred $\Delta H_0$ in the isotropic lightcones increase significantly with a larger scatter in the scaling relation (injected or not). The $\MT$ relation has the smallest scatter and thus shows a better constraining power than $\LT$ and $\YT$. Lower $\sigma_\mathrm{intr}$ relations will play an important role in future studies.

By carefully accounting for cosmic variance and statistical noise, the tension of the detected M21 anisotropy and $\Lambda$CDM is substantially reduced but remains significant. This work highlights the necessity of simulations as an integral part of interpreting statistical significance and isolating genuine cosmological anomalies from sample variance. As next-generation datasets from surveys such as eRASS become available, extending this analysis with real data alongside simulated cluster catalogues will be crucial for corroborating these results and refining our understanding of the observed cosmological anisotropies.

\begin{acknowledgements}
{We thank the anonymous referee for their valuable comments that helped us improve our paper.} K.M. acknowledges support in the form of the X-ray Oort Fellowship at Leiden
Observatory. This work used the DiRAC@Durham facility managed by the Institute
for Computational Cosmology on behalf of the STFC DiRAC HPC Facility
(www.dirac.ac.uk). The equipment was funded by BEIS capital funding via STFC
capital grants ST/P002293/1, ST/R002371/1 and ST/S002502/1, Durham University
and STFC operations grant ST/R000832/1. DiRAC is part of the National
e-Infrastructure.
\end{acknowledgements}

%
%

\bibliographystyle{aa}
\bibliography{ClusterAnisotropy}

\begin{appendix}

\onecolumn

\section{Two 1 Gpc runtime lightcones}

\begin{figure}[h!]
    \centering
    \includegraphics[width=0.5\linewidth]{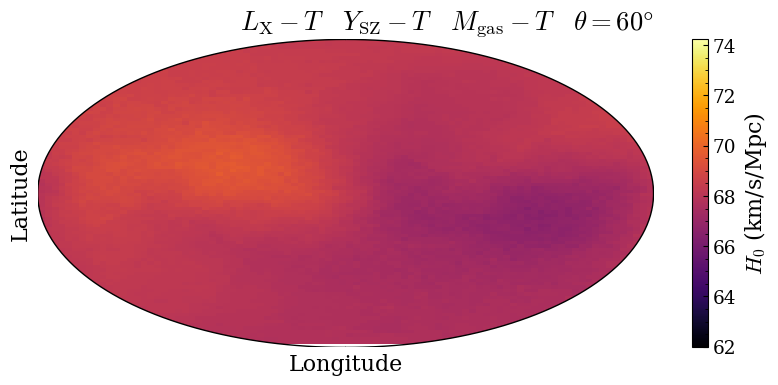}
    \includegraphics[width=0.5\linewidth]{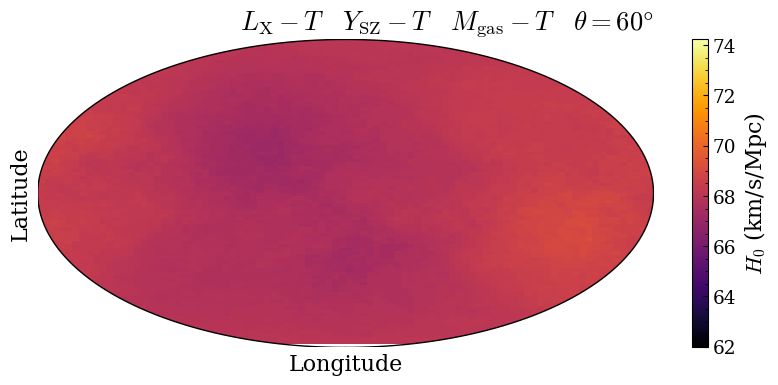}
    \caption{$H_0$ maps from the M21 method for the two runtime lightcones in 
    the $\SI{1.0}{Gpc}$ fiducial run (L1\_m9), using the combined $\LT$, $\YT$, 
    and $\MT$ relations. Lightcone 1 (top) and lightcone 2 (bottom).}
    \label{fig:appendix-h0-combined-lc0lc1}
\end{figure}

Applying the same analysis to the two runtime lightcones from the L1\_m9 run, we  find a maximum dipole of $3.15 \pm 0.78\%$ at $4.02\sigma$ in lightcone 1, and  $2.50 \pm 0.71\%$ at $3.55\sigma$ in lightcone 2, using the M21 method  (see Fig.~\ref{fig:appendix-h0-combined-lc0lc1}). These runtime lightcones are  consistent with the 1728 post-processed lightcones analysed in the main text.  No detection exceeds $3\sigma$ from the main probability distribution of the  1728 samples. This support the robustness of our post-processed lightcones.

\FloatBarrier

\section{Details on the fiducial mass cut}\label{app:mass-cut}

\begin{figure}[h!]
    \centering
    \includegraphics[width=0.5\linewidth]{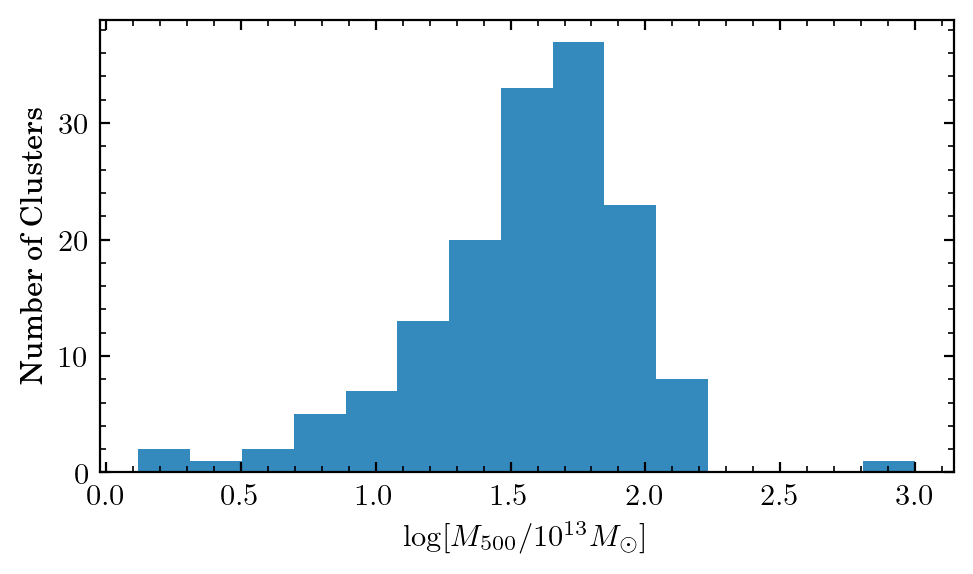}
    \caption{Mass distribution of eRASS1 clusters satisfying $f_\mathrm{X,\,0.2{-}2.3\,\mathrm{keV}} > \SI{5e-12}{erg\,s^{-1}\,cm^{-2}}$. None fall below the fiducial mass cut of $10^{13}\,\mathrm{M}_\odot$.}
    \label{fig:mass-cut-erass}
\end{figure}

To reduce computational cost, we have applied a fiducial mass cut of $M_\mathrm{500c} > 10^{13}\,\mathrm{M}_\odot$ (Sect.~\ref{sec:lightcone-construction}). We verify here that this selection does not exclude clusters that would otherwise satisfy the flux threshold. Using the publicly available eRASS1 catalogue \citep{bulbul_srgerosita_2024}, we apply the same flux selection of $f_\mathrm{X,\,0.2{-}2.3\,\mathrm{keV}} > \SI{5e-12}{erg\,s^{-1}\,cm^{-2}}$. The resulting mass distribution is shown in Fig.~\ref{fig:mass-cut-erass}. Out of 12247 clusters detected by eROSITA, 152 meet the flux threshold, none of which have $M_\mathrm{500c} < 10^{13}\,\mathrm{M}_\odot$.

On the simulation side, applying the same flux threshold to our lightcones results in, on average, fewer than five clusters within the range $10^{13}\,\mathrm{M}_\odot < M_\mathrm{500c} < 10^{13.5}\,\mathrm{M}_\odot$ (see Fig.~\ref{fig:lightcones-distribution}). These systems lie at the low-mass tail. Thus, even if the mass cut excludes a few rare clusters that could surpass the flux threshold, they form an extremely small subset and do not contribute meaningfully. We conclude that the mass cut does not introduce a selection bias.

\FloatBarrier
\section{Correlation of results using different relations: angular
separation}\label{app:angular-separation}

\begin{figure}[h!]
    \centering
    \begin{subfigure}[b]{0.5\linewidth}
        \centering
        \includegraphics[width=1.0\linewidth]{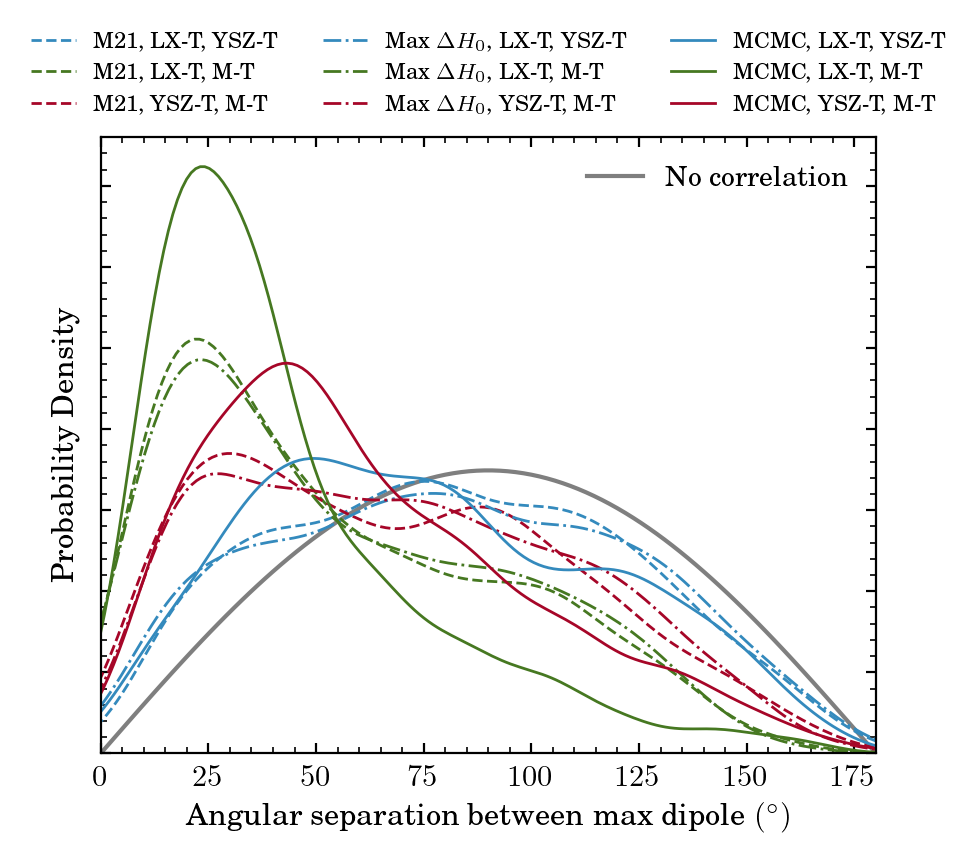}
        \subcaption{No scatter, all three relation pairs}
    \end{subfigure}
    \begin{subfigure}[b]{0.5\linewidth}
        \centering
        \includegraphics[width=0.85\linewidth]{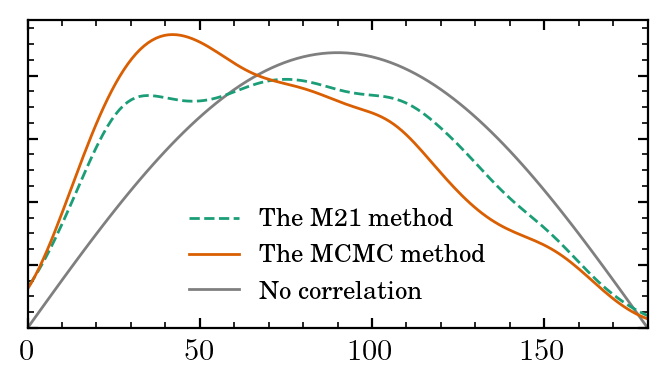}
        \subcaption{Matched scatter, $\LT$ and $\YT$}
    \end{subfigure}
    \caption{Angular separation between the best-fit dipoles derived from different relations, without (top) and with (bottom) injected scatter. Linestyles indicate statistical methods: solid for MCMC; dashed and dash-dotted for the M21 method, differing slightly in implementation. Dashed lines define the dipole by highest significance (our default for the M21 method); dash-dotted lines by highest amplitude. The two give nearly identical results. The grey line indicates the expected distribution for uncorrelated dipoles.}
    \label{fig:angular-separation}
\end{figure}

See Fig.~\ref{fig:angular-separation}. We compute the angular separation on the sphere between best-fit dipole directions derived from different scaling relations. Comparing with Fig.~\ref{fig:scatter-correlation}, one observes that greater scatter correlation results in closer dipole alignment. Relations with higher Pearson correlation coefficients yield smaller angular separations in our lightcones. Even for the best-aligned pair, $\LT$ and $\MT$ (via MCMC), only 77\% of lightcones have separation angle $\theta < 60\degree$. This drops to 56\% for $\YT$ and $\MT$ and 39\% for $\LT$ and $\YT$. The figure also confirms consistency between the two statistical methods: the choice of relation affects the dipole direction more strongly than the method used. Comparing the bottom panel and the blue curves, no significant change is seen when scatter is injected. This is expected—no intrinsic signal is present, and the injected scatter has no cross-relation correlation between $\LT$ and $\YT$ (see Appendix~\ref{app:scatter-correlation}).

For uncorrelated directions, the expected angular separation distribution is 
$P(\theta) = \frac{1}{2}\sin{\theta}$. This follows from uniform sampling on 
the unit sphere: the probability of a second point lying between angles 
$\theta$ and $\theta + d\theta$ from a fixed point is proportional to the area 
of the corresponding spherical ring, which scales as $2\pi \sin{\theta} d\theta$. 
The normalisation factor $1/2$ ensures unit probability.

\FloatBarrier

\section{Sample comparison between FLAMINGO and M21}\label{app:sample-comparison}

\subsection{X-ray concentration}\label{app:xray-concentration}

\begin{figure}[h!]
    \centering
    \includegraphics[width=0.5\linewidth]{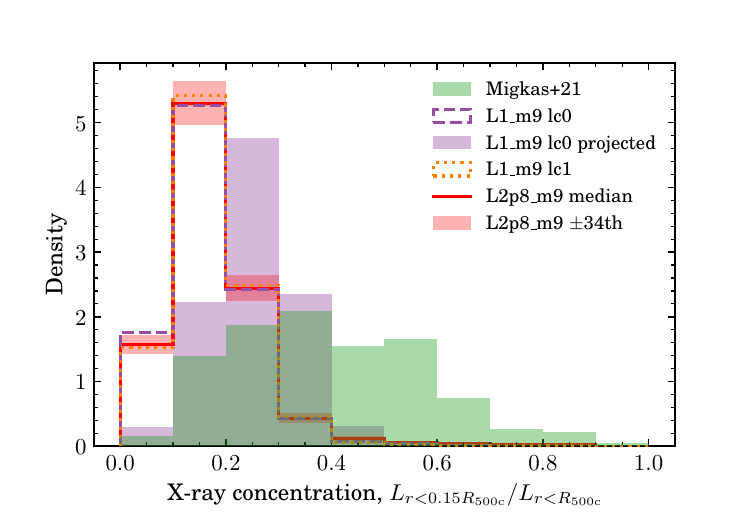}
    \caption{X-ray concentration distribution of 1728 $(\SI{2.8}{Gpc})^3$
    lightcones (red), two $\SI{1.0}{Gpc}$ runtime lightcones (purple dashed,
    orange dotted), and the M21 sample (green). Purple bars show the 2D
    projected concentration of a $\SI{1.0}{Gpc}$ lightcone.}
    \label{fig:xray-concentration}
\end{figure}

The distribution of $c_{\mathrm{X}}$, defined as the fraction of emission coming from the central 15\% of the cluster's radius ($L_{r<0.15R_\mathrm{500c}}/ L_\mathrm{r<R_\mathrm{500c}}$), is an easily estimated metric that helps us identify potentially relaxed clusters with high central emission. Large $c_{\mathrm{X}}$ values often correlate with cool-core presence in clusters \citep{andrade-santos_fraction_2017, kugel_flamingo_2024}. $c_{\mathrm{X}}$ exhibits the most prominent difference between FLAMINGO and the M21 sample (see Fig.~\ref{fig:xray-concentration}). \citet{kugel_flamingo_2024} investigated the distribution of the same quantity and found the same behaviour when the projection effect was not taken into account. However, Fig.~\ref{fig:xray-concentration} shows that the difference is much reduced, but not eliminated, by the projection effect. The projected $L_\mathrm{X}$ are calculated by summing particles inside cylindrical regions instead of spherical ones. The FLAMINGO samples' projected $c_{\mathrm{X}}$ is centred at $\sim 0.25$ whereas the M21 sample's distribution is centred at $c_{\mathrm{X}}\sim 0.35$. There are few FLAMINGO clusters with $c_{\text{X}}>0.35$. This might be caused by an under-representation of high X-ray concentration clusters in FLAMINGO in addition to non-detections of diffuse, low $c_{\mathrm{X}}$ clusters in X-ray flux-limited samples, especially close to the flux limit. Moreover, highly disturbed clusters (expected to return lower $c_{\text{X}}$ values) were removed from the M20/M21 sample. Nevertheless, cluster samples selected through their SZ signal \citep[e.g.][]{lovisari_x-ray_2017} also show a relatively higher $c_{\text{X}}$ distribution than FLAMINGO. In any case, further investigation of this discrepancy is needed.

\subsection{Cluster properties comparison}\label{app:cluster-properties-comparison}

Fig.~\ref{fig:lightcones-distribution} summarises the distributions of $f_\mathrm{X}$, $L_\mathrm{X}$, $T_\mathrm{Cl}$, $M_\mathrm{500c}$, $Y_\mathrm{SZ}$, and $M_\mathrm{gas}$ for FLAMINGO, the runtime lightcones, and M21. Simulated samples are selected as the top 313 X-ray concentrated clusters to match the number of available M21 entries; results are robust to changing this number to $250\dash400$. Simulated $L_\mathrm{X}$ values are in the original SOAP band ($0.2\text{--}\SI{2.3}{keV}$; see Sect.~\ref{sec:sample-Lx}), with M21 values converted accordingly. Agreement between FLAMINGO and M21 is generally good across properties, though $T_\mathrm{Cl}$ is systematically lower in FLAMINGO (see Sect.~\ref{sec:sample-T}). In addition, $L_\mathrm{X}$, $f_\mathrm{X}$, and $Y_\mathrm{SZ}$ in FLAMINGO also tend to lie slightly on the lower end, which may be partly due to an under-representation of high-redshift clusters. The runtime lightcones are consistent with the simulated samples.

\begin{figure*}[h!]
    \centering
    \includegraphics[width=0.49\linewidth]{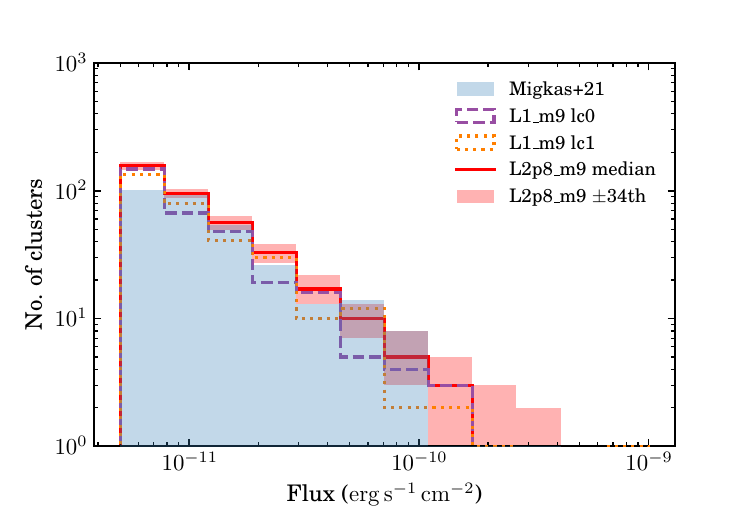}
    \includegraphics[width=0.49\linewidth]{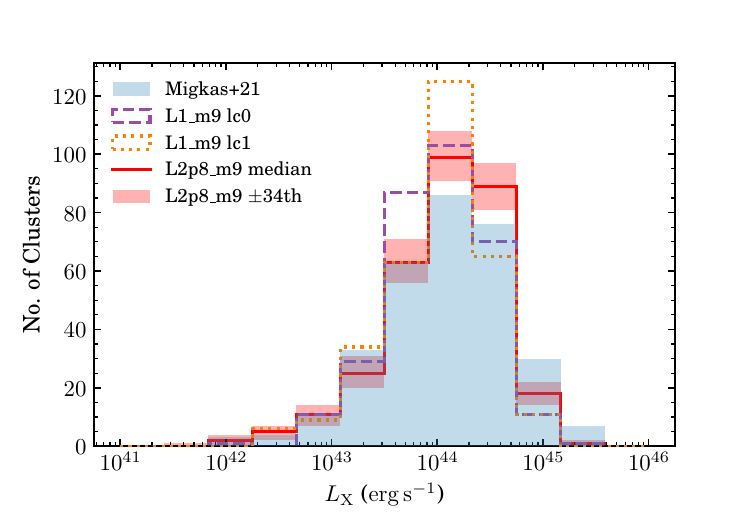}
    \caption{Distribution of various cluster properties of the simulated
    samples, runtime lightcones, and the M21 sample, where available (continued next page).}
    \label{fig:lightcones-distribution}
\end{figure*}%
\begin{figure*}[h!]\ContinuedFloat
    \centering
    \includegraphics[width=0.49\linewidth]{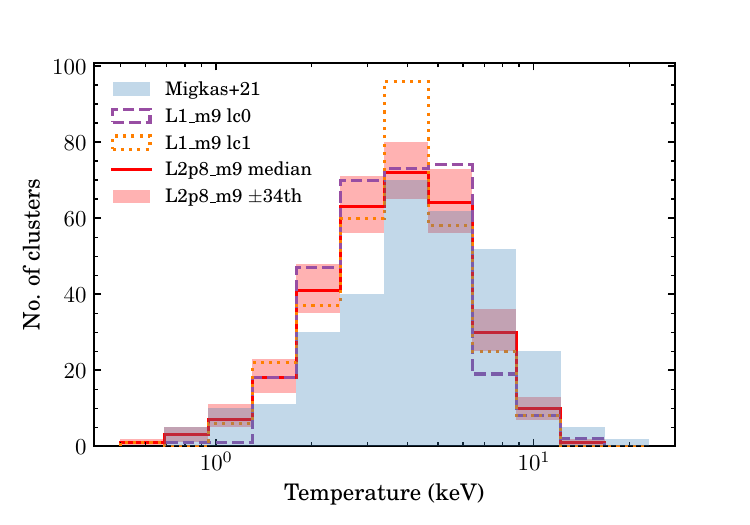}
    \includegraphics[width=0.49\linewidth]{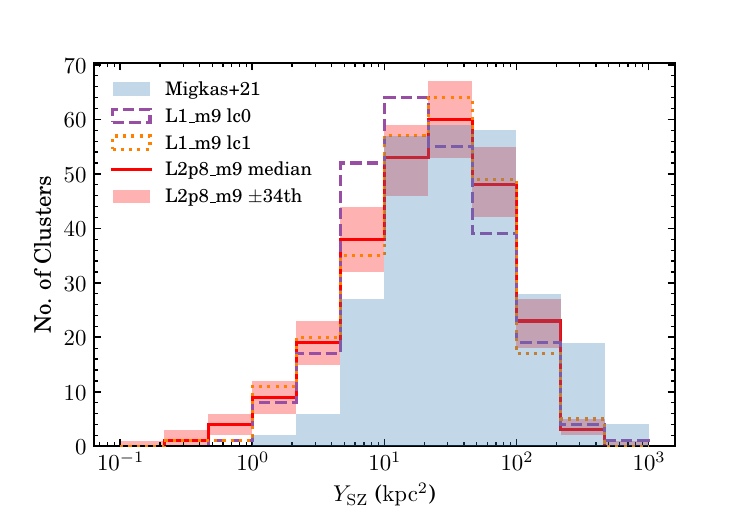}
    \includegraphics[width=0.49\linewidth]{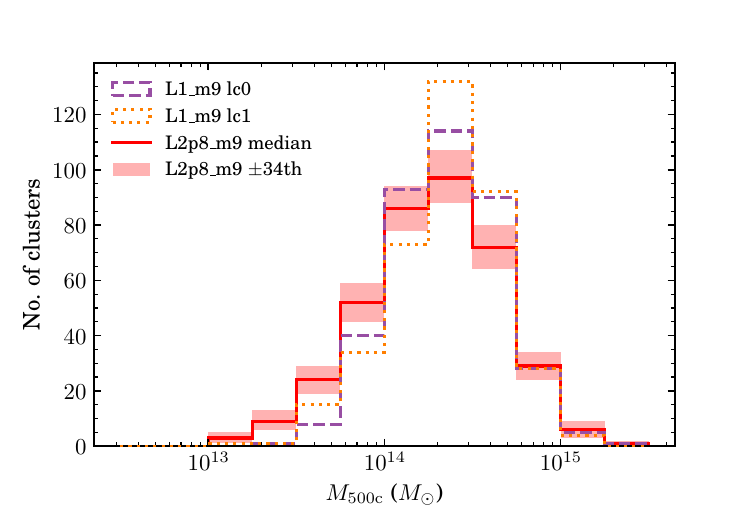}
    \includegraphics[width=0.49\linewidth]{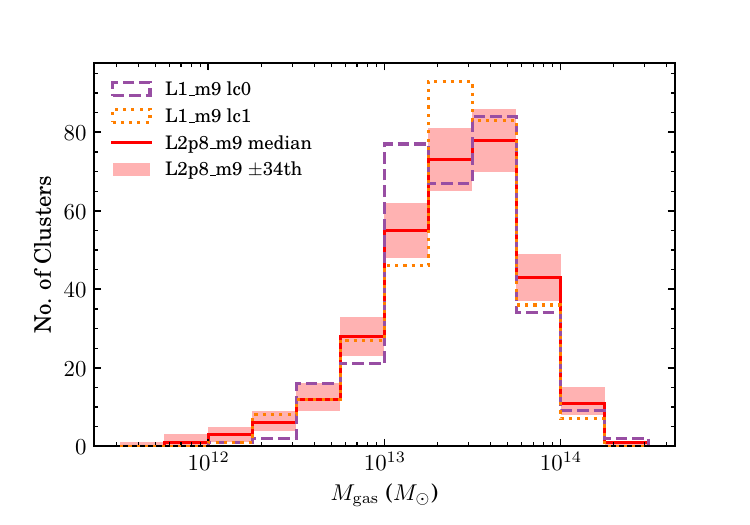}
    \caption{Continued.}
    \label{fig:lightcones-distribution}
\end{figure*}

\FloatBarrier

\section{Scatter correlation between the relations}\label{app:scatter-correlation}

\begin{figure*}
    \centering
    \includegraphics[width=0.3\linewidth]{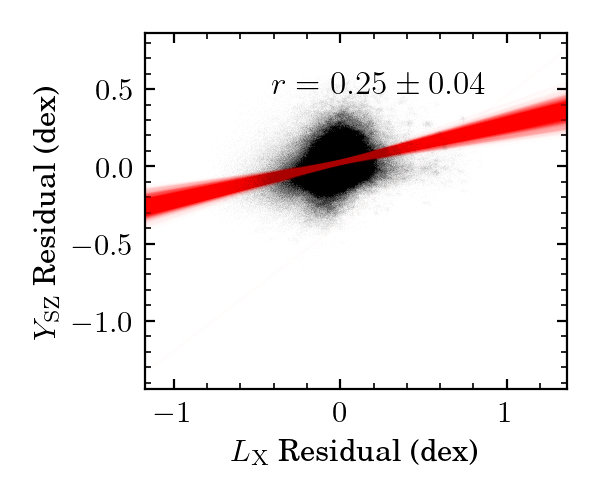}
    \includegraphics[width=0.3\linewidth]{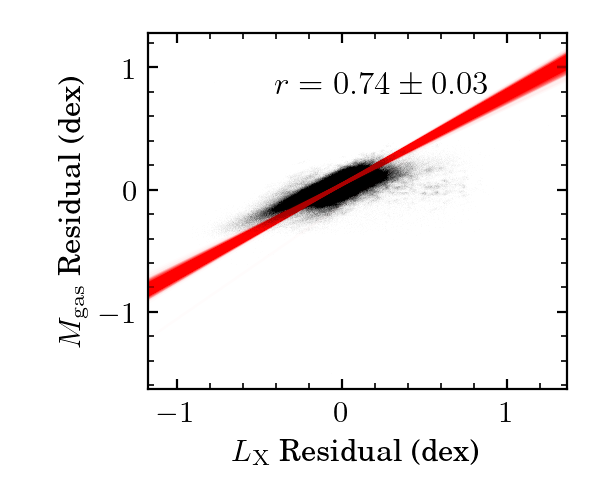}
    \includegraphics[width=0.3\linewidth]{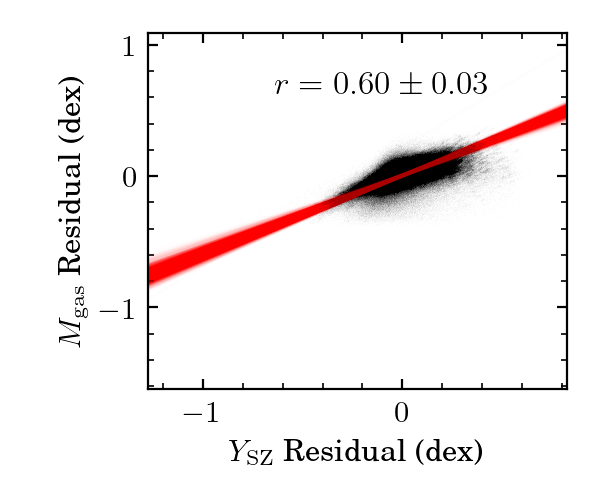}
    \caption{Correlation of scatter between $\LT$ and $\YT$ (left), $\LT$ and
    $\MT$ (middle) and $\YT$ and $\MT$ (right). Black dots are individual
    clusters. Red lines represent the linear fit of the lightcones. All of the 1728
    lightcones are overlaid. $r$ is Pearson's correlation coefficient.}
    \label{fig:scatter-correlation}
\end{figure*}

Scatter is computed as $\log Y - \log Y_\text{fit}$, where $\log Y_\text{fit} = B \log X / C_X + \log A + \log C_Y - \gamma_{YX} \log E(z)$. We evaluate scatter for $\LT$, $\YT$, and $\MT$, and fit a linear relation between each pair across all 1728 lightcones. The resulting scatter points and fits are shown in Fig.~\ref{fig:scatter-correlation}, with Pearson's correlation coefficient $r$ labelled in each panel. The uncertainty is estimated from the standard deviation of the fit results across all lightcones. 

Using $T_\text{Cl}$ as the common variable for comparison with M21, we find $r = 0.25 \pm 0.04$ between $\LT$ and $\YT$ in FLAMINGO. This is significantly lower than the $r = 0.668 \pm 0.089$ found in M21. To assess the effect of injected scatter, we repeated the analysis with artificial scatter applied. Since the injection procedure does not induce cross-correlation, the result remains similar: $r = 0.31 \pm 0.03$. \citet{nagarajan_weak-lensing_2019} reported $r = 0.47^{+0.24}_{-0.35}$ for the $L_\mathrm{X}$–$M_{500}$ and $Y_\mathrm{SZ}$–$M_{500}$ relations at fixed mass. Such correlation could arise from baryonic physics linking X-ray and SZ emission, which may not be fully captured in our modelling.

In contrast, we find stronger correlations between $L_\mathrm{X}$ and $M_\mathrm{gas}$, and between $Y_\mathrm{SZ}$ and $M_\mathrm{gas}$ at fixed $T$, with $r = 0.74 \pm 0.03$ and $r = 0.60 \pm 0.03$, respectively. Future studies using $M_\mathrm{gas}$-based relations should be cautious when combining observables or constructing mock samples.

\FloatBarrier

\section{Correlation of anisotropy using the M21 or MCMC method}\label{app:m21-vs-mcmc}

\begin{figure*}[h!]
    \centering
    \begin{subfigure}[b]{0.3\linewidth}
        \centering
        \includegraphics[width=\linewidth]{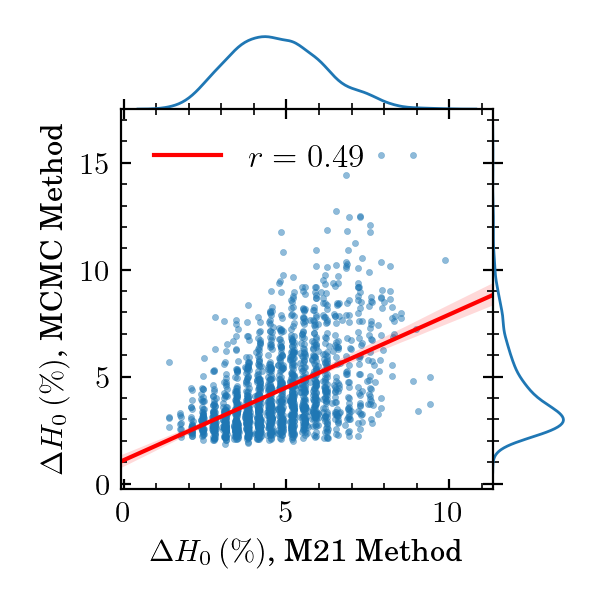}
        \subcaption{$\LT$, $\Delta H_0$}
    \end{subfigure}
    \begin{subfigure}[b]{0.3\linewidth}
        \centering
        \includegraphics[width=\linewidth]{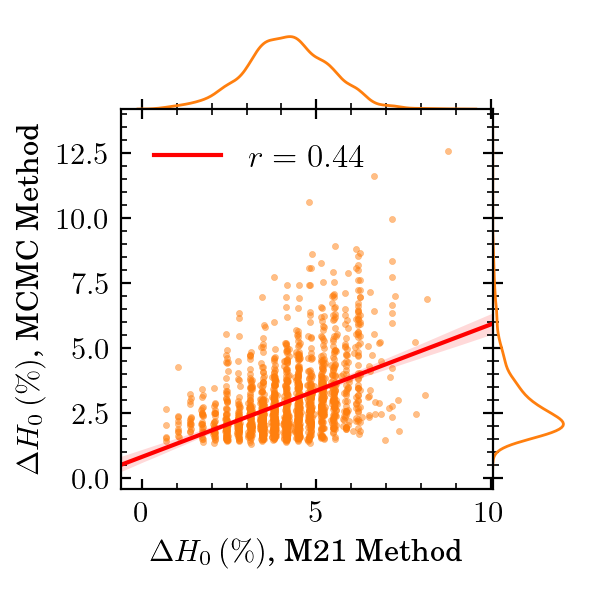}
        \subcaption{$\YT$, $\Delta H_0$}
    \end{subfigure}
    \begin{subfigure}[b]{0.3\linewidth}
        \centering
        \includegraphics[width=\linewidth]{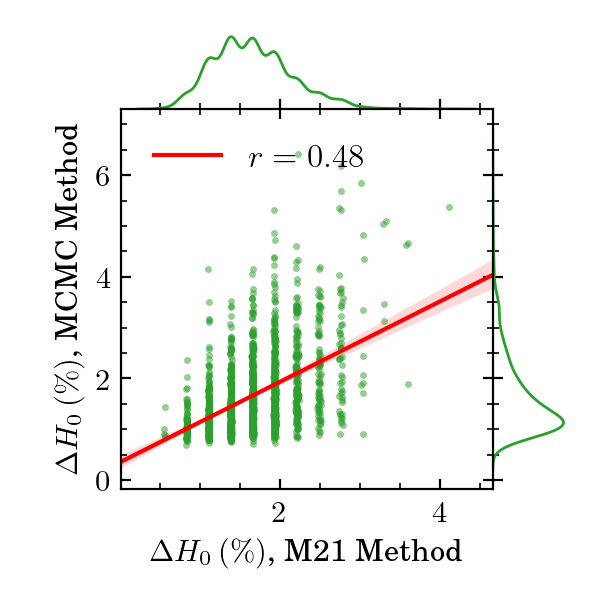}
        \subcaption{$\MT$, $\Delta H_0$}
    \end{subfigure}
    \begin{subfigure}[b]{0.3\linewidth}
        \centering
        \includegraphics[width=\linewidth]{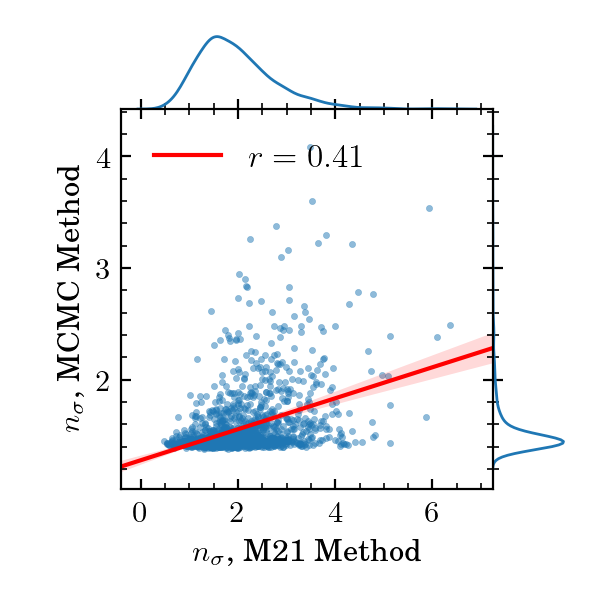}
        \subcaption{$\LT$, $n_\sigma$}
    \end{subfigure}
    \begin{subfigure}[b]{0.3\linewidth}
        \centering
        \includegraphics[width=\linewidth]{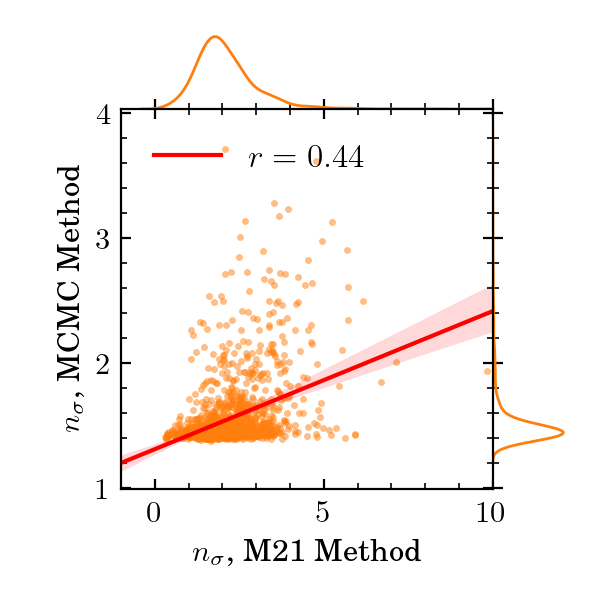}
        \subcaption{$\YT$, $n_\sigma$}
    \end{subfigure}
    \begin{subfigure}[b]{0.3\linewidth}
        \centering
        \includegraphics[width=\linewidth]{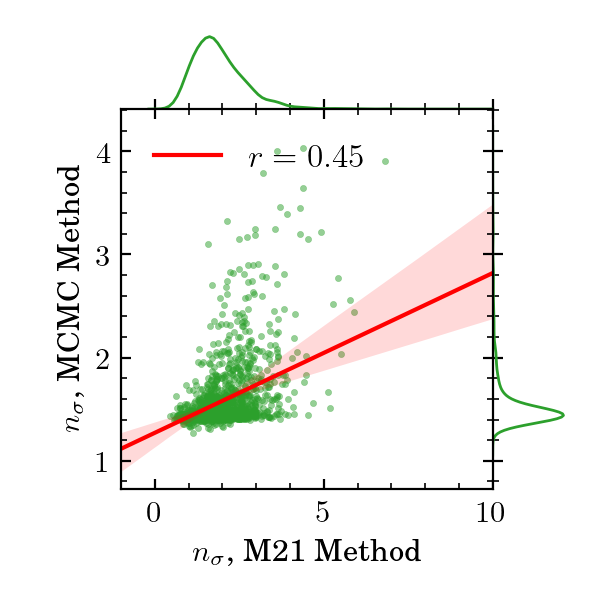}
        \subcaption{$\MT$, $n_\sigma$}
    \end{subfigure} 
    \caption{Comparing $\Delta H_0$ (top) and $n_\sigma$ (bottom) predicted by the MCMC method versus the M21 method, using the $\LT$ (left), $\YT$ (middle), or $\MT$ (right) relation. Each data point represents the maximum dipole $\Delta H_0$ or $n_\sigma$ on a lightcone. The red line shows linear fit with error. $r$ represents the Pearson correlation coefficient.}
    \label{fig:m21-vs-mcmc}
\end{figure*}

The correlation of amplitude $\Delta H_0$ and statistical significance $n_\sigma$ between the M21 method and the MCMC approach is shown in Fig.~\ref{fig:m21-vs-mcmc}. A uniform mild correlation of $r \sim 0.4\dash 0.5$ is observed across all six comparisons, covering the three scaling relations and both anisotropy metrics. This test is carried out without injected scatter. The scatter around the one-to-one line suggests that the two methods do not fully agree on the strength of anisotropy for individual realisations, but they tend to rank lightcones similarly in terms of dipole strength. This level of agreement supports the overall consistency between methods, while highlighting that method-dependent systematics could still influence the detailed interpretation of individual detections.

\FloatBarrier

\section{Extreme value statistics}\label{app:about-evs}

\begin{figure*}[h!]
    \centering
    \begin{subfigure}[b]{0.24\linewidth}
        \includegraphics[width=\linewidth]{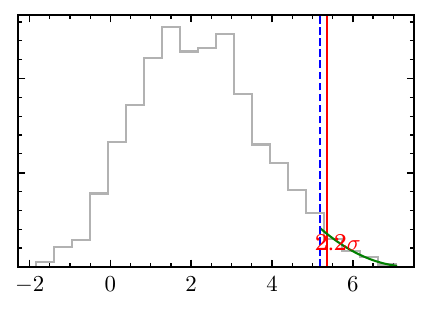}
        \subcaption{$\LT$, M21}
    \end{subfigure}
    \begin{subfigure}[b]{0.24\linewidth}
        \includegraphics[width=\linewidth]{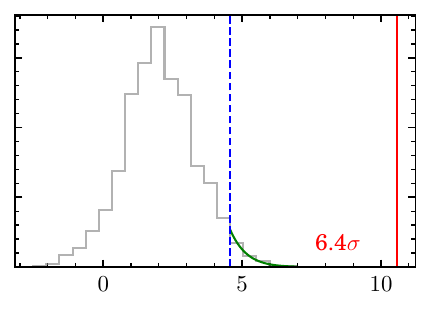}
        \subcaption{$\YT$, M21}
        \label{fig:evs-m21-YT}
    \end{subfigure}
    \begin{subfigure}[b]{0.24\linewidth}
        \includegraphics[width=\linewidth]{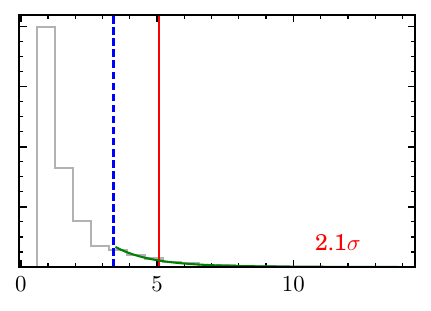}
        \subcaption{$\LT$, MCMC}
    \end{subfigure}
    \begin{subfigure}[b]{0.24\linewidth}
        \includegraphics[width=\linewidth]{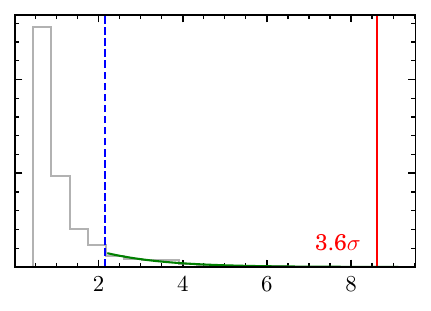}
        \subcaption{$\YT$, MCMC}
    \end{subfigure}
    \begin{subfigure}[b]{0.24\linewidth}
        \includegraphics[width=\linewidth]{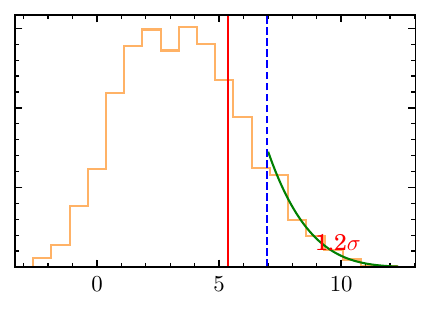}
        \subcaption{$\LT$ with scatter, M21}
    \end{subfigure}
    \begin{subfigure}[b]{0.24\linewidth}
        \includegraphics[width=\linewidth]{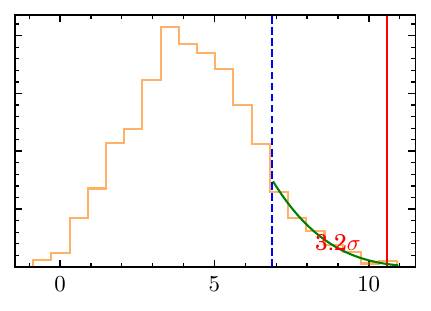}
        \subcaption{$\YT$ with scatter, M21}
        \label{fig:evs-m21-YT-scatter}
    \end{subfigure}
    \begin{subfigure}[b]{0.24\linewidth}
        \includegraphics[width=\linewidth]{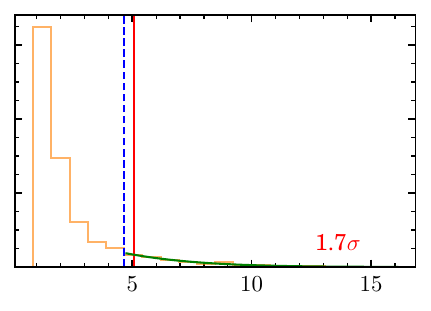}
        \subcaption{$\LT$ with scatter, MCMC}
    \end{subfigure}
    \begin{subfigure}[b]{0.24\linewidth}
        \includegraphics[width=\linewidth]{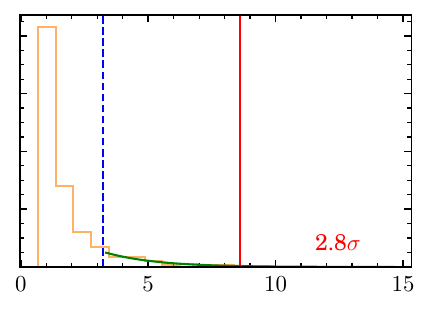}
        \subcaption{$\YT$ with scatter, MCMC}
    \end{subfigure}
    \caption{$H_0$ anisotropy EVS probability distributions for individual scaling 
relations. Legend as in Fig.~\ref{fig:evs-all-joint}. Corresponding $p$-values 
are listed in Table~\ref{tab:h0-$p$-value-summary}.}
    \label{fig:evs-LT-and-YT}
\end{figure*}

\begin{figure*}[h!]
    \centering
    \begin{subfigure}[b]{0.48\linewidth}
        \centering
        \includegraphics[width=\linewidth]{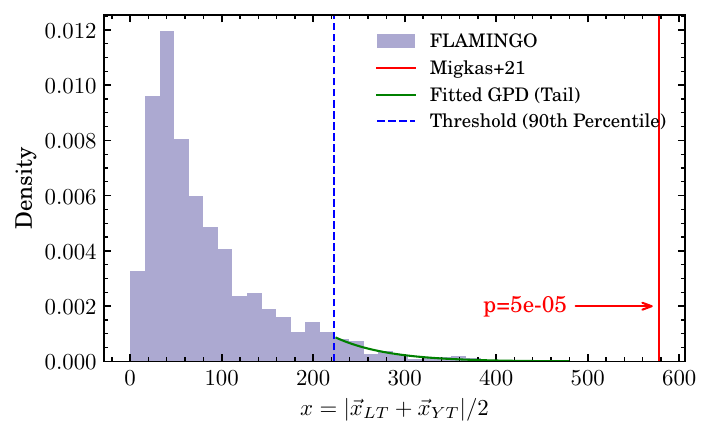}
        \subcaption{$z<0.07$}
    \end{subfigure}
    \begin{subfigure}[b]{0.48\linewidth}
        \centering
        \includegraphics[width=\linewidth]{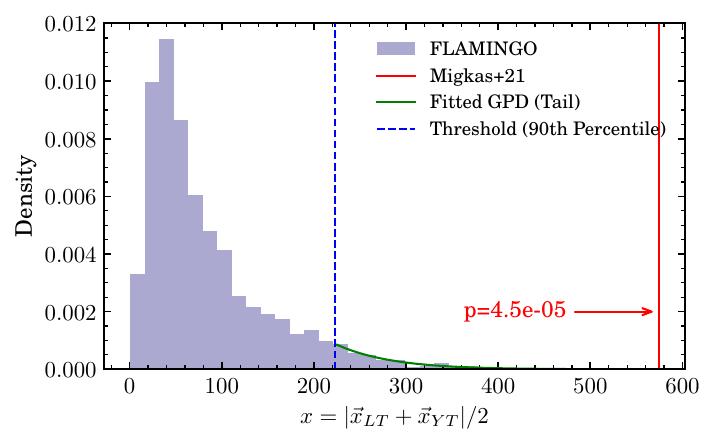}
        \subcaption{$z<0.10$}
    \end{subfigure}
    \begin{subfigure}[b]{0.48\linewidth}
        \centering
        \includegraphics[width=\linewidth]{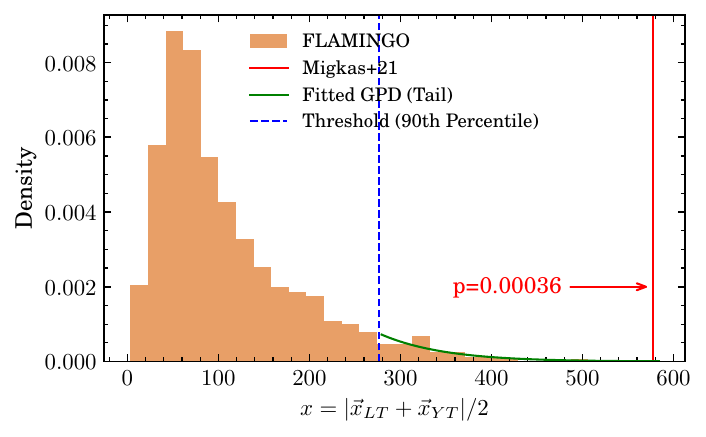}
        \subcaption{$z<0.07$, matched scatter}
    \end{subfigure}
    \begin{subfigure}[b]{0.48\linewidth}
        \centering
        \includegraphics[width=\linewidth]{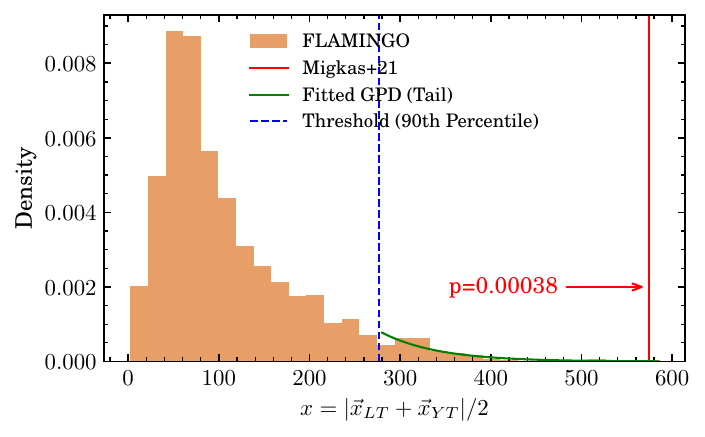}
        \subcaption{$z<0.10$, matched scatter}
    \end{subfigure}
    \caption{Joint EVS analyses for bulk flows, before (top) and after (bottom) matching scatter. Legend as in Fig.~\ref{fig:evs-all-joint}. Corresponding  $p$-values are listed in Table~\ref{tab:h0-$p$-value-summary}.}
    \label{fig:evs-bulk-flow-joint}
\end{figure*}

\begin{figure*}[h!]
    \centering
    \begin{subfigure}[b]{0.24\linewidth}
        \includegraphics[width=\linewidth]{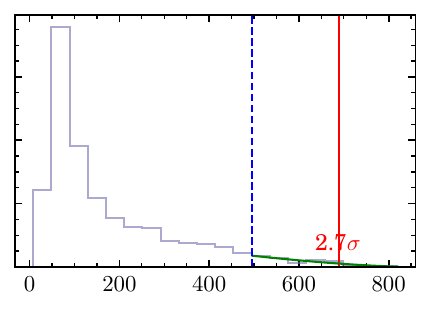}
        \subcaption{$\LT$, $z<0.07$}
    \end{subfigure}
    \begin{subfigure}[b]{0.24\linewidth}
        \includegraphics[width=\linewidth]{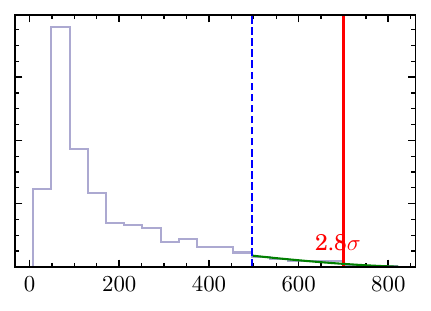}
        \subcaption{$\LT$, $z<0.10$}
    \end{subfigure}
    \begin{subfigure}[b]{0.24\linewidth}
        \includegraphics[width=\linewidth]{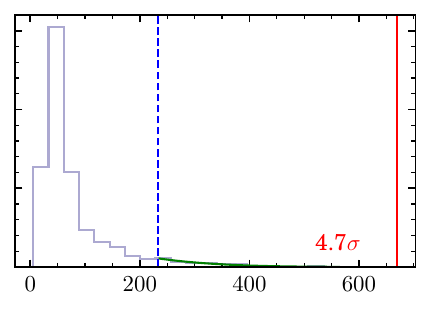}
        \subcaption{$\YT$, $z<0.07$}
    \end{subfigure}
    \begin{subfigure}[b]{0.24\linewidth}
        \includegraphics[width=\linewidth]{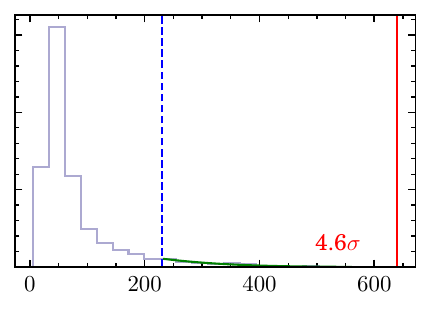}
        \subcaption{$\YT$, $z<0.10$}
    \end{subfigure}
    \begin{subfigure}[b]{0.24\linewidth}
        \includegraphics[width=\linewidth]{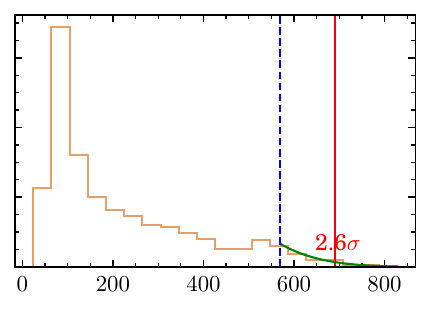}
        \subcaption{$\LT$ with scatter, $z<0.07$}
    \end{subfigure}
    \begin{subfigure}[b]{0.24\linewidth}
        \includegraphics[width=\linewidth]{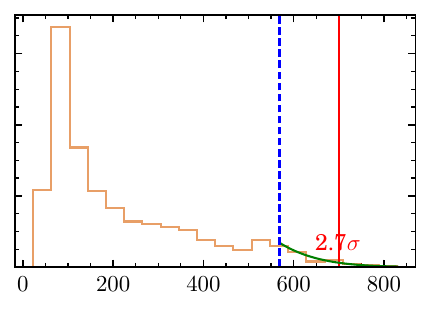}
        \subcaption{$\LT$ with scatter, $z<0.10$}
    \end{subfigure}
    \begin{subfigure}[b]{0.24\linewidth}
        \includegraphics[width=\linewidth]{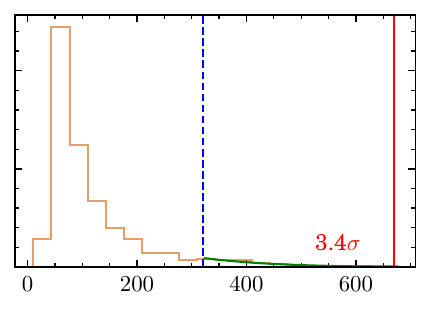}
        \subcaption{$\YT$ with scatter, $z<0.07$}
    \end{subfigure}
    \begin{subfigure}[b]{0.24\linewidth}
        \includegraphics[width=\linewidth]{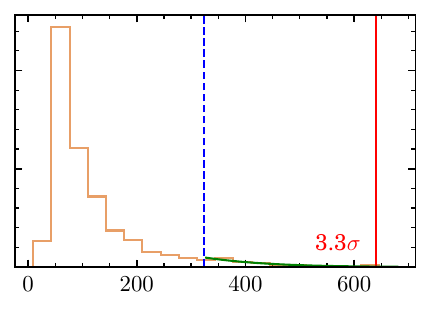}
        \subcaption{$\YT$ with scatter, $z<0.10$}
    \end{subfigure}
    \caption{Bulk flow EVS probability distributions for individual scaling 
    relations. Legend as in Fig.~\ref{fig:evs-all-joint}. Corresponding results 
    are summarised in Table~\ref{tab:results-bf}.}
    \label{fig:evs-bulk-flow-per-relation}
\end{figure*}

Here we show the EVS estimations for other than the joint $H_0$ (Fig.~\ref{fig:evs-all-joint}). Every entry of EVS $p$-values in Table~\ref{tab:h0-$p$-value-summary} and \ref{tab:results-bf} is accompanied by a figure. For Table~\ref{tab:h0-$p$-value-summary} the per relation EVS distributions are presented in Fig.~\ref{fig:evs-LT-and-YT}. For the $\LT$, M21 or MCMC method entries the reference point from M21 is $(\Delta H_0, n_\sigma)=(8.7\%, 2.8\sigma)$ or $(8.7\%. 2.4\sigma)$. For $\YT$ we compare to $(14\%, 4.1\sigma)$ with the M21 method or $(14\%, 2.6\sigma)$ with MCMC. Same as in Fig.~\ref{fig:evs-all-joint} the tension is greatly relieved by increased scatter, where the projected $x$ value (Sect.~\ref{sec:method-probability}) is shifted closer to the M21 points and with increased scatter (see Fig.~\ref{fig:evs-m21-YT-scatter} in comparison to Fig.~\ref{fig:evs-m21-YT}, for example).

The joint EVS analyses for bulk flow (Table~\ref{tab:results-bf}) are summarised in Fig.~\ref{fig:evs-bulk-flow-joint}, and results for individual relations in  Fig.~\ref{fig:evs-bulk-flow-per-relation}. As with $H_0$, the projected distributions  shift upward when scatter is added, though the effect is visibly weaker. In particular, the tension in $\LT$ remains largely unchanged.

Other than $x = \Delta H_0 - \Delta H_0 / n_\sigma$, several other projection functions were tested for the $H_0$ matched scatter analysis (see Table~\ref{tab:evs-projections}). We see that the choice of projection has a noticeable effect on the results. The choice of projection has a clear impact on the resulting $p$-values, underscoring the need for a physically motivated and consistent definition. Our adopted metric, the lower one-sigma bound, was chosen to incorporate both anisotropy amplitude and uncertainty in a single conservative estimate. It ensures that a modest but well-constrained dipole (e.g. $5 \pm 1\%$) is treated comparably to a larger but poorly constrained one (e.g. $15 \pm 11\%$), avoiding overinterpretation of noisy outliers.

\begin{table*}[h!]
    \centering
    \footnotesize
    \caption{Effect of different projection functions on the EVS $p$-values.}
    \begin{tabular}{ccccccc}
    \toprule
        Projection & M21 $\LT$ & M21 $\YT$ & M21 joint & MCMC $\LT$ & MCMC $\YT$
        & MCMC joint \\
    \midrule
        $\Delta H_0 - \Delta H_0 / n_\sigma$ & 
        $0.231\,(1.2\sigma)$ & $0.00129\,(3.22\sigma)$ & $0.00118\,(3.24\sigma)$ &
        $0.0848\,(1.72\sigma)$ & $0.00487\,(2.82\sigma)$ & $0.0205\,(2.32\sigma)$
        \\
        $\Delta H_0 / 4 + n_\sigma$ & 
        $0.232\,(1.2\sigma)$ & $0.0205\,(2.32\sigma)$ & $0.00619\,(2.74\sigma)$ &
        $0.0877\,(1.71\sigma)$ & $0.00576\,(2.76\sigma)$ & $0.00079\,(3.36\sigma)$
        \\
        $\Delta H_0 / 2 + n_\sigma$ &
        $0.250\,(1.15\sigma)$ & $0.0139\,(2.46\sigma)$ & $0.00486\,(2.82\sigma)$ & 
        $0.114\,(1.58\sigma)$ & $0.00447\,(2.84\sigma)$ & $0.00491\,(2.81\sigma)$
        \\
        $\Delta H_0 + \Delta H_0 / n_\sigma$ &
        $0.394\,(0.85\sigma)$ & $0.000297\,(3.62\sigma)$ & $0.00235\,(3.04\sigma)$ & $0.237\,(1.18\sigma)$ & $0.00251\,(3.02\sigma)$ & $0.142\,(1.47\sigma)$
        \\
    \bottomrule
    \end{tabular}
    \label{tab:evs-projections}
\end{table*}

\end{appendix}

\end{document}